\documentclass[fleqn,usenatbib]{mnras}

\usepackage{newtxtext,newtxmath}

\usepackage[T1]{fontenc}

\DeclareRobustCommand{\VAN}[3]{#2}
\let\VANthebibliography\thebibliography
\def\thebibliography{\DeclareRobustCommand{\VAN}[3]{##3}\VANthebibliography}

\usepackage{graphicx}	
\usepackage{amsmath}	
\usepackage[flushleft]{threeparttable}

\usepackage{soul}

\title[Observations of UID 30901]{Two years of optical and NIR observations of the superluminous supernova UID 30901 discovered by the UltraVISTA SN survey}

\author[Hueichapan et al.]{
E. D. Hueichapan,$^{1,2}$\thanks{E-mail: emilio.hueichapan@mail.udp.cl}
C. Contreras,$^{3}$
R. Cartier,$^{1,4}$
P. Lira,$^{2}$
P. Sanchez-Saez,$^{5}$
B. Milvang-Jensen,$^{6}$
\newauthor
J. P. U. Fynbo,$^{6}$  J.\,P. Anderson$^{7}$ and M. Hamuy $^{8,9}$
\\
$^{1}$Cerro Tololo Inter-American Observatory, NSF's National Optical-Infrared Astronomy Research Laboratory, Casilla 603, La Serena, Chile\\
$^{2}$Departamento de Astronom\'ia, Universidad de Chile, Camino el Observatorio 1515, Las Condes, Santiago, Casilla 36-D, Chile.\\
$^{3}$Las Campanas Observatory, Carnegie Observatories, Casilla 60, La Serena, Chile\\
$^{4}$Gemini Observatory, NSF’s National Optical-Infrared Astronomy Research Laboratory, Casilla 603, La Serena, Chile\\
$^{5}$European Southern Observatory, Karl-Schwarzschild Str. 2, D-85748 Garching bei M\"unchen, Germany\\
$^{6}$Cosmic Dawn Center (DAWN), Copenhagen, Denmark; Niels Bohr Institute, University of Copenhagen, Copenhagen, Denmark\\
$^{7}$European Southern Observatory, Alonso de Córdova 3107, Casilla 19, Santiago, Chile\\
$^{8}$Fundación Chilena de Astronomía, Santiago, Chile\\
$^{9}$Hagler Institute for Advanced Studies, Texas A\&M University, Texas, USA\\
}

\date{Accepted XXX. Received YYY; in original form ZZZ}

\pubyear{2022}

\begin{document}
\label{firstpage}
\pagerange{\pageref{firstpage}--\pageref{lastpage}}
\maketitle

\begin{abstract} 
We present deep optical and near-infrared photometry of UID 30901, a superluminous supernova (SLSN) discovered during the UltraVISTA survey. The observations were obtained with VIRCAM ($YJHK_{s}$) mounted on the VISTA telescope, DECam ($griz$) on the Blanco telescope, and SUBARU Hyper Suprime-Cam (HSC; $grizy$). These multi-band observations comprise +700 days making UID 30901 one of the best photometrically followed SLSNe to date.
The host galaxy of UID 30901 is detected in a deep HST F814W image with an AB magnitude of $27.3 \pm 0.2$. While no spectra exist for the SN or its host galaxy, we perform our analysis assuming $z = 0.37$, based on the photometric redshift of a possible host galaxy found at a projected distance of 7 kpc.
Fitting a blackbody to the observations, the radius, temperature, and bolometric light curve are computed. We find a maximum bolometric luminosity of $5.4 \pm 0.34 \times 10^{43}$ erg s$^{-1}$. A flattening in the light curve beyond 600 days is observed and several possible causes are discussed. We find the observations to clearly favour a SLSN type I, and plausible power sources such as the radioactive decay of $^{56}$Ni and the spin-down of a magnetar are compared to the data. We find that the magnetar model yields a good fit to the observations with the following parameters: a magnetic field $B = 1.4 \pm 0.3 \times 10^{14}$ G, spin period of $P = 6.0 \pm 0.1 $\,ms and ejecta mass $M_{ej} = 11.9^{+4.8}_{-6.4} M_{\odot}$. 
\end{abstract}




\section{Introduction}

Two decades ago a new kind of extremely bright stellar explosion
was uncovered, now known as superluminous supernovae (SLSNe). 
These objects are rare and can often reach absolute magnitudes
of $M\sim -21$ at maximum light \citep{GalYam19}. It was soon
realised that some of these SNe exhibit hydrogen in their spectra,
while others do not, giving as a result Type II and Type I SLSNe 
sub-classes \citep{galyam12} respectively, in analogy with normal
luminosity SNe \citep[see][]{filippenko97}.

Hydrogen-poor SLSNe are generally characterised by a very blue, nearly 
featureless optical spectrum, a lack of hydrogen lines and the 
display of characteristic \ion{O}{ii} absorption lines near maximum
light. The presence of \ion{O}{ii} absorption lines is a 
signature of the high temperature and ionisation state of the SN 
ejecta at early times \citep[e.g,][]{mazzali16}. A few weeks after
maximum, when the ejecta has cooled enough, their spectra become
similar to SNe Ic \citep{pastorello10} or to broad-line Ic SNe
\citep[Ic-Bl][]{liu17}. 

After a few hundred days, their nebular spectra are 
dominated by intermediate mass elements, resembling SNe 
Ic-BL \citep{milisavljevic13, nicholl16b, jerkstrand17, nicholl19}.
Recently, it has been noted that hydrogen poor SLSNe span a wide 
range in peak luminosities ($-22 \lesssim M_g \lesssim -20$ mag), 
overlapping with the Ic-BL SN class \citep{decia18}. 
However, while some degree of similarity exists between SLSNe-i and Ic/Ic-BL, 
\citet{decia18} and \citet{quimby18} have recently claimed that
hydrogen-poor SLSNe are both photometrically and spectroscopically
a distinct SN class. 

Hydrogen-rich or Type II SLSNe are less common objects than SLSNe-I, and are characterised by a distinctive H$\alpha$ emission line. There is large diversity in the observational signatures of hydrogen-rich SLSNe.  In this class, we find objects such as SN\,2006gy \citep{smith07, Ofek07} which are presumed to be powered by strong SN ejecta circumstellar medium (CSM) interaction. These kinds of SLSNe display narrow and broad H$\alpha$ emission components, which are characteristic of Type IIn SNe \citep{schlegel90}. Objects such as SN\,2006gy are frequently considered the bright end of the Type IIn class. Among SLSNe-II, we can also find objects such as SN\,2008es \citep{miller09,gezari09}, displaying a very blue and featureless continuum at early times, no \ion{O}{ii} lines or narrow lines in the spectra, but developing a strong and broad dominant H$\alpha$ feature after a few days.
The interaction between the SN ejecta with a dense CSM is less evident in objects like SN\,2008es, but maybe the main power source. The diversity of the SLSNe-II class was further characterised through two objects reported by \citet{inserra18}.

There is an interesting subset of objects initially classified as SLSNe-I,
such as iPTF16bad and iPTF13ehe \citep{yan15,yan17}, which show broad H$\alpha$ 
emission after maximum light and signatures of SN ejecta-CSM interaction. These 
objects suggest ejections of hydrogen-rich material shortly before the SN 
explosion. These mass ejections may be more common in SLSNe-I than currently 
detected, although radio \citep{nicholl16b} and X-ray \citep{margutti18} 
observations favour low density environments similar to SNe Ic. However, due to 
the high redshifts at which SLSNe are frequently found, it is often difficult to 
secure high signal-to-noise observations in X-rays, radio, or in the optical/NIR 
at  late times to place strong constraints on the environment surrounding the SN.

In the last decade, several researchers have proposed diverse physical
mechanisms to power the extreme luminosity of SLSNe. They can be summarised
as: 1) the interaction between a fast SN ejecta and a dense CSM that
transforms the ejecta kinetic energy into thermal energy, 2) the radioactive
decay of several solar masses of $^{56}$Ni synthesised during the SN
explosion and 3) the power injection from a central engine such as fallback 
accretion onto a black hole \citep{dexter13} or a rapidly rotating neutron star 
with a strong magnetic field -a magnetar- formed during the core collapse 
\citep{woosley10, kasen10}.

Most SLSNe-II exhibit hydrogen signatures and light curves
with maximum luminosity, duration, shape and decline
rate that seem to be well explained within the context of ejecta-CSM 
interaction models. An example of this is the analytical ejecta-CSM 
interaction model implemented by \citet{Chatzopoulos13} which
can reproduce most of the diversity exhibited by the light curves
of this type of event. This model assumes that the progenitor star
is surrounded by a CSM shell described by a power law density profile
$\rho_{\mathrm{CSM}}=qr^{- s}$, where 0 or 2 are physically 
motivated values for $s$. The first value describes a shell of constant
density and the latter a steady-wind. However, we have to consider
that simple analytical models although powerful to obtain an estimate
of CSM properties, are not able to capture the full complexity of the 
ejecta-CSM interaction \citep[see e.g.,][]{moriya18}. For example, the
pre-SN mass loss history can be more complex resulting in CSM clumps
or shells, with non-spherical distributions, resulting in a diverse
set of SN light curves and spectra, depending on the viewing angle which may,
in part, explain the observational diversity observed in SLSNe-II.

In the case of SLSN-I, however, the power source is less clear. One possibility 
is that their extreme luminosity can be due to ejecta-CSM interaction
as in SLSNe-II, but in this case considering a fast SN ejecta with a 
hydrogen and helium poor CSM \citep[see][for a review]{moriya18}. 
The handful objects initially classified as SLSNe-I at early times, 
but later showing broad H$\alpha$ emission and other signatures of 
CSM-ejecta interaction \citep{yan15, yan17}, may bridge the gap between
SLSNe-I and SLSNe-II populations. However, several challenges remain for 
the interaction model to explain the SLSN-I population. For example, 
whether it is possible for a massive star to expel large enough quantities of 
hydrogen-free material to power SLSNe light curves through
an ejecta-CSM interaction. Another important question for this
scenario is whether the spectral features observed in SLSNe-I can
be produced by ejecta-CSM interaction. 
It is important to consider that several days after maximum light
SLSNe-I spectra resemble non-interacting stripped envelope
core collapse SNe (Ic and Ic-BL), suggesting that
strong interaction does not play a role in the formation of
the spectral features in SLSNe-I.

A different route to power SLSNe is through the radioactive decay of several 
solar masses of $^{56}$Ni synthesised in a Pair Instability SN 
\citep{galyam09, kasen11}. In the exceptional instance of stars born in low 
metallicity environments with a main-sequence mass in the range 140-260 
$M_{\odot}$ \citep{heger02}, an instability due to the production of 
positron-electron pairs can lead to a thermonuclear explosion 
\citep{barkat67,ravaky67}, known as a PISN.

PISNe are predicted to ejecta several solar masses of 
$^{56}$Ni. Its radioactive decay products and other iron-peak elements
should produce: 1) strong blanketing below $\sim4000$ \AA\ shifting 
the peak of the spectral energy distribution (SED) towards redder 
wavelengths \citep{dessart12}, and 2) a nebular spectrum
dominated by emission lines from iron-peak elements 
\citep{dessart12, jerkstrand16,mazzali19}. These signatures of PISN are 
in stark contrast with the characteristics displayed by SLSNe-I, 
which: 1) exhibit a very blue SED before and around maximum light, 
peaking below $4000$ \AA, and 2) have nebular spectra dominated
by emission lines from intermediate-mass elements
\citep{milisavljevic13, jerkstrand17, nicholl19, mazzali19}. 

A popular model that can reproduce the high luminosities observed in SLSNe-I, 
invokes the spin-down of a fast rotating and highly magnetic neutron star, that 
energizes the SN ejecta \citep{woosley10, kasen10}. Using a small sample of 
well-observed hydrogen poor SLSNe, \citet{inserra13} discarded the decay of 
$^{56}$Ni as the power source for their objects and instead showed that the 
energy injection from the spin-down of a magnetar can successfully reproduce the 
complete light curve evolution, including their flattening at late times. More 
recently, \citet{nicholl17} presented a sophisticated version of the magnetar 
model built-in the Modular Open Source Fitter for Transients 
\citep[\sc MOSFiT]{guillochon18}, which they fit to a large sample of hydrogen 
poor SLSNe collected from the literature to show that this model can 
successfully explain the light curve evolution of this SN class. 

The characterisation of SLSNe host galaxies also plays an important role in 
constraining their possible progenitors. Several studies have shown that 
SLSNe-I have a strong preference for low mass dwarf galaxies 
($M_{\mathrm{stellar}} < 10^{9} M_{\odot}$) \citep{neill11, lunnan14, leloudas15, angus16, perley16, schulze18}, with metallicity usually below 0.5\, Z$_{\sun}$ 
\citep{lunnan14, leloudas15, perley16, schulze18} and high specific star 
formation rates \citep{neill11, lunnan14, perley16}. On the other hand, 
SLSNe-II seems to explode in a wider range of environments, from the extreme
environments typical of SLSNe-I to those of normal luminosity core collapse SNe 
\citep{leloudas15, perley16, schulze18}. 

In this paper, we present and analyse $YJHK_s$ photometry of 
UID 30901 spanning more than 700 days of observations from the UltraVISTA survey.
UltraVISTA is a deep near-infrared (NIR) galaxy survey, running from 2009 to 
2016. Although not originally designed as a transient survey, it is well suited 
for finding transients thanks to its depth ($\sim 23.5$ mag in individual 
images), multi-band observations and high-cadence. 

We complement these data with $griz$ photometry from archival images from the
Dark Energy Camera (DECam) \citep{flaugher15} and $grizy$ photometry from the
SUBARU Hyper Suprime-Cam (HSC) \citep{aihara19}. The SLSN was found several
years after the explosion and there are no spectra available of the explosive
event nor for its possible very faint host galaxies. To perform our analysis we
use the photometric redshifts from the COSMOS2015 catalog
\citep{laigle16}. The full $grizyYJHK_{s}$ set of photometry, makes the UID
30901 one of the best observed SLSNe to date.

This paper is organised as follows: in Section \ref{sec:observations}
we overview the instruments used and present UID 30901 photometry. In Section 
\ref{sec:analysis} we determine the host galaxy properties, the light curve
and derive the main physical parameters for our SLSN. We discuss
the different power sources and apply physical models to fit our dataset
in Section \ref{sec:discussion}, and finally present our conclusions
in Section \ref{sec:summary}.

All magnitudes in this paper are expressed in the AB system. We adopt a
$\Lambda$CDM cosmology with the Hubble constant $H_{0} = 70$ km\,s$^{-1}$\,Mpc$^{-1}$,
total dark matter density $\Omega_{M} = 0.3$ and dark energy density
$\Omega_{\Lambda}=0.7$.

\section{Observations} \label{sec:observations}

We present more than 700 days of optical and NIR observations in nine filters
of UID 30901, a SN discovered as part of the UltraVISTA SN survey. 
The UltraVISTA SN survey is a NIR time-domain survey to search for 
SNe and Kilonovae (KNe) on the time resolved data obtained by UltraVISTA 
\citep{mccracken12}. The search for transients was performed two years after the 
end of the data collection, therefore we do not have spectra for most of the 
transients discovered by the survey, including 
UID 30901. We will describe the UltraVISTA SN survey in detail in a future publication.
All the NIR $YJHK_{s}$ photometry presented here was obtained as part 
of the UltraVISTA survey itself, and the optical ($grizy$) data correspond
to archival data obtained with the Dark Energy Camera (DECam)
mounted on the Blanco telescope at Cerro Tololo in Chile, and from
Hyper Suprime-Cam (HSC) mounted on the Subaru telescope located
in Mauna Kea, Hawaii. 

\begin{figure}
    \includegraphics[width=8cm]{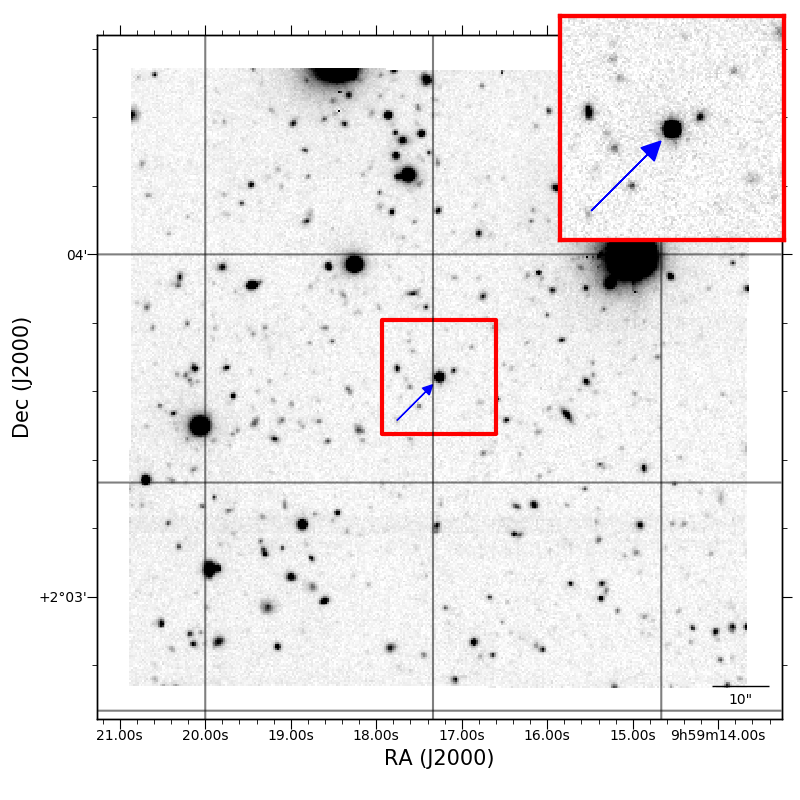}
    \caption{Finding Chart for UID 30901. Subaru $r$-band
    of UID 30901 taken on March, 2014.}
    \label{fig:fchart}
\end{figure}

UID 30901 was discovered in the UltraVISTA data and the first
detection epoch corresponds to the same night on images obtained with 
DECam and UltraVISTA on March 17, 2014, with magnitudes of 
$g = 21.91 \pm 0.12$ and $K_{s} = 22.81 \pm 0.14$. The last 
non-detection previous to the discovery was on March 12th on images 
obtained by UltraVISTA in the $H-$band to a depth of $H = 23.3 \pm 0.1$. 
The SN is located in the COSMOS field at 
$\alpha = 09^{\mathrm{h}}59^{\mathrm{m}}17\fs25$, 
$\delta=+02\degr03\arcmin38\farcs5$ (see Figure \ref{fig:fchart}),
close to two faint galaxies described in Section \ref{sec:host}.
We measured PSF photometry on the background subtracted images for this
object in the optical and in the NIR for the following two years after
discovery, until March 2016. A galactic reddening correction was applied
to the light curves of UID 30901 given by $E(B-V) = 0.0172$ 
\citep{Schlafly11} following the \citet{Cardelli89} law with $R_{V} = 3.1$.

\subsection{NIR photometry}

We present the time-domain $YJHK_{s}$ photometry obtained as part of 
the UltraVISTA survey, carried out between the 14th of December 2009 and
the 29th of June 2016. UltraVISTA used VIRCAM \citep{dalton06}, a 
wide-field NIR camera mounted on the Cassegrain focus of the 4.1\,m
VISTA telescope \citep{emerson06, emerson10} at Paranal Observatory.
VIRCAM consists of 16 $2048 \times 2048$ Raytheon VIRGO HgCdTe arrays
with a mean pixel scale of 0\farcs34 pixel$^{-1}$. 
Even though the UltraVISTA survey was aimed to explore distant
galaxies, their high cadence, the multi-wavelength coverage, 
the depth of the images ($\sim 23$ mag) and the extension of the
survey make it optimal for the search for transients.

We worked with processed images, which correspond to image stacks of OB blocks 
of typical total exposure times of 0.5 hr or 1 hr. The processed images contain
good astrometric information in their headers, so no further astrometric 
refinement was required.  At the location of UID 30901, there is no significant 
host galaxy emission. However, to discover transients and remove their host 
galaxies a generic template subtraction strategy was applied to UltraVISTA data. 
The subtraction templates were constructed from images obtained in 2009 and 
2010, several years before the SN explosion in the case of UID 30901. 
To perform the alignment between the template image and the images to be 
subtracted we used {\sc swarp} \citep{bertin02}, and {\sc hotpants}\footnote{https://github.com/acbecker/hotpants} to do the image 
subtraction. {\sc hotpants} uses an algorithm from \citet{alard00} for the 
creation and application of a spatially varying convolution kernel.

Photometry was performed using a custom PSF fitting code calibrated
against an UltraVISTA catalog of stars in the AB system. The UltraVISTA
$YJHK_{s}$ photometry is presented in Figure \ref{fig:opnir_lc} and
summarised in Table \ref{tab:nir} in the Appendix. 

\begin{figure*}
    \includegraphics[width=12cm]{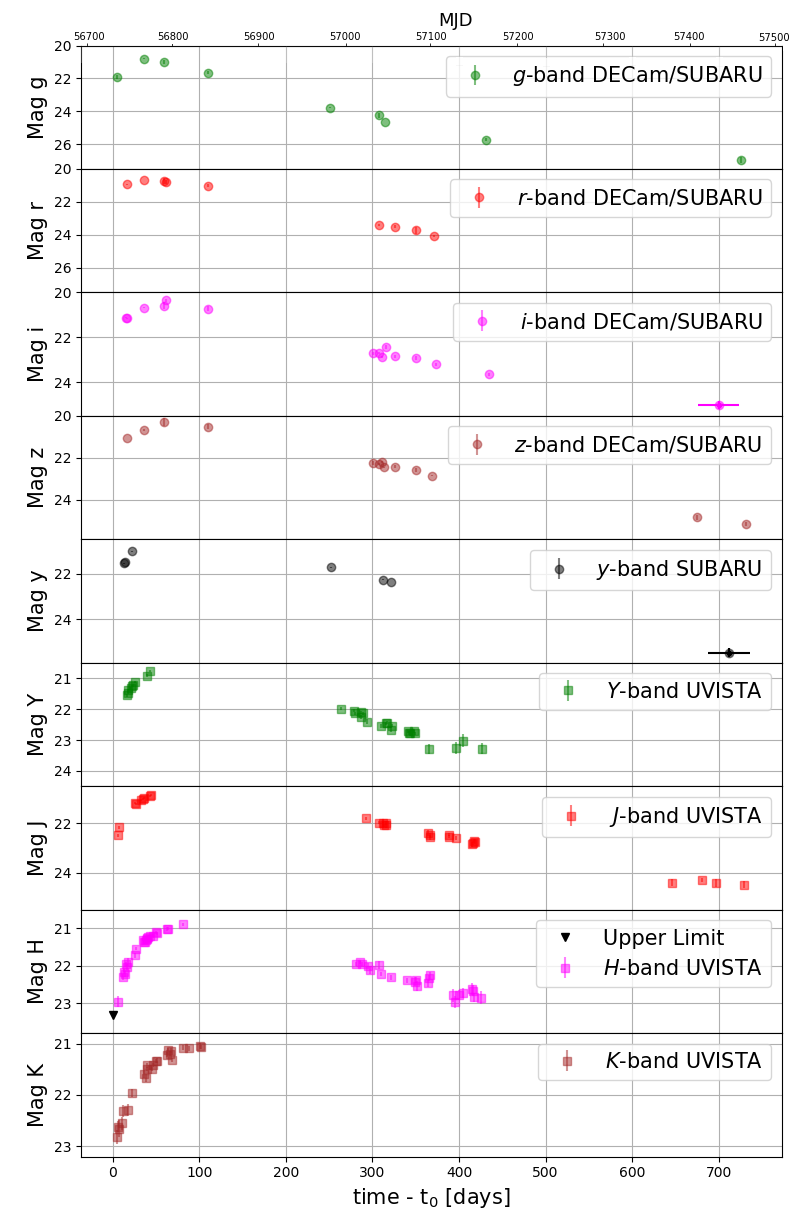}
    \caption{Optical and NIR photometry of the UID 30901}
    \label{fig:opnir_lc}
\end{figure*}

\subsection{Optical photometry}

In addition to the UltraVISTA NIR photometry, we present optical
photometry computed from archival images obtained with the DECam
instrument \citep{flaugher15} mounted on the Blanco 4-m telescope,
and with the HSC \citep{aihara19} mounted on the 8.2\,m Subaru
telescope. The DECam images were downloaded using the 
archival online tool from the NOIRLab Database\footnote{https://astroarchive.noao.edu/portal/search/}. Deep
HSC $grizy$ images were downloaded from the HSC online interface
\footnote{https://hsc-release.mtk.nao.ac.jp/doc/index.php/tools-2/}, 
comprising images from 2014 until 2017. From these images we obtained 
27 $griz$ and 25 $grizy$ individual measurements for DECam and HSC,
respectively. 

The $grizy$ PSF photometry was measured relative to a series of seven isolated 
stars close to the SN. The $griz$-band photometry of 
the stars was obtained from the Sloan Digital Sky Survey
\citep{sdssdr13} database, while the $y$-band photometry was 
downloaded using HSC online tools. The optical photometry of the SN 
is summarised in Table \ref{tab:optical} of the Appendix and 
presented in Figure \ref{fig:opnir_lc}. The photometry of the 
field stars is presented in Table \ref{tab:sdss} of the Appendix.

\section{Analysis} \label{sec:analysis}

\begin{figure*}
\begin{center}
\includegraphics[width=18cm]{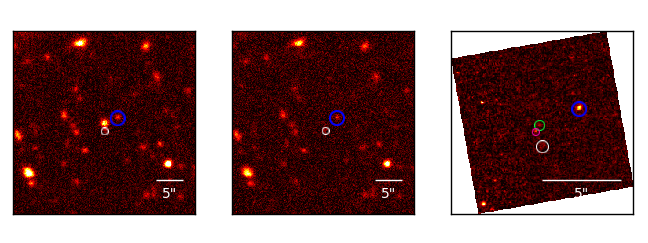}\\
\includegraphics[width=12cm]{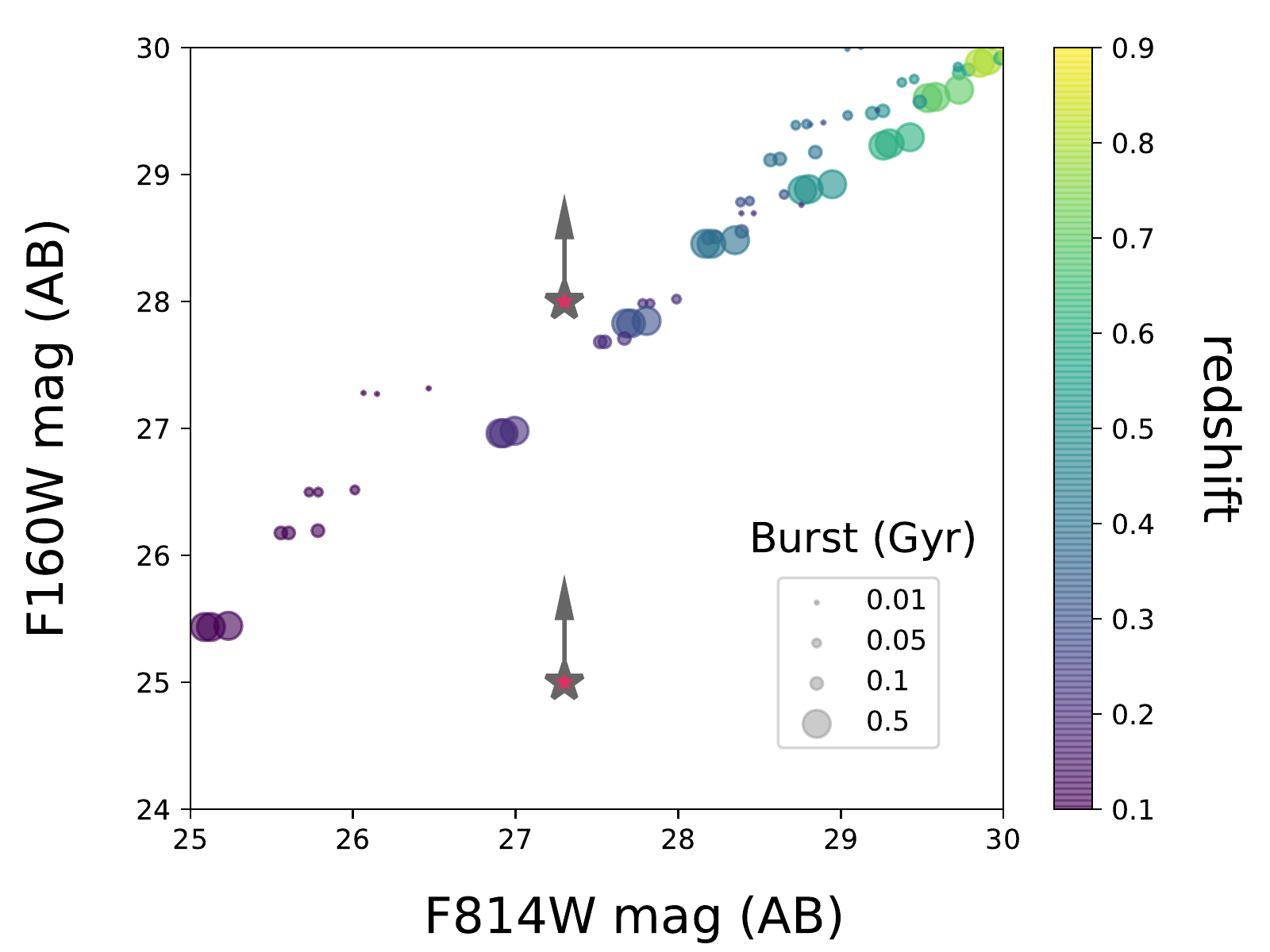}
\caption{Finding charts for UID 30901, its host galaxy and possible galaxy
  companions. North is up and East is left. \textit{Left:} SUBARU g-band with
  UID 30901 in the center. \textit{Middle:} SUBARU g-band after last detection
  of the SN. \textit{Right:} F814W image from HST WFC where the host is
  detected. White and blue circles mark the position of companion galaxies A
  and B, respectively.  The green circle marks the most likely host which
  corresponds to the position of UID 30901 while the magenta circle presents a
  single detection of a non reported SN from 2004. Bottom: Synthetic F814W and
  F160W magnitudes obtained for the host galaxy using parameters from
  \protect\cite{perley16}. Colours represent the redshift of the galaxy and the
  size of the circles represent the duration of the burst. The photometry of
  the host is also included with two upper limits shown for the F160W filter
  (see text for details).}
\label{fig:hosts}
\end{center}
\end{figure*}

In this Section, we analyse the data of UID 30901. This includes
determining the properties of the host galaxy, measuring the explosion date, 
characterising the light curves, deriving blackbody 
parameters during the early evolution and the fitting of models 
to the full light curves to determine the most likely power engine. 

\subsection{Host Galaxy} \label{sec:host}

SLSNe have been found generally in low-mass faint galaxies \citep{schulze18},
which makes their hosts increasingly difficult to detect with increasing 
redshifts.
An HST WFC image obtained on April 8th, 2004 using the F814W filter reveals a
very faint object at the exact position of UID 30901, which we identify as its 
host (see Figure \ref{fig:hosts}) and another point-like source that is likely to
correspond to another transient, possibly another SN exploding in the same host 
for which no further observations are available. Photometry gives a magnitude of 
$27.5 \pm 0.3$ for the host and $26.9 \pm 0.1$ for the unidentified SN. HST 
NICMOS observations of the same region of the sky taken on September 9th, 2009 using the F160W filter, however, fail to show any detection. Upper limits of $\sim 28$ and  $\sim 25$ magnitudes were obtained for the host assuming a point-like and galaxy-like shape, respectively. The galaxy used to model the extended shape is, however, clearly more extended than the actual host, as can be seen in the F814W image (its identification is given below), and therefore we estimate that a more realistic upper limit is found somewhere between these two extreme values.

A recent study by \cite{orum2020} reveals that about 50\%\ of dwarf galaxies
hosting SLSNe are found in crowded regions, independent of the redshift of the
SN. Since we do not know the redshift of the host associated to UID 30901, we
searched other extended extra-galactic sources with a projected distance close
to the SN site as these could be companions of the host of UID 30901.

COSMOS2015 \citep{laigle16} is a photometric redshift catalogue including deep 
imaging from UltraVISTA DR2 \citep{mccracken12}, SUBARU (Suprime-Cam and Hyper 
Suprime-Cam) and Spitzer \citep[SPLASH catalog, see][Section 2]{steinhardt14} 
among other telescope-instrument combinations from previous surveys of the 
$\sim$ 2 degrees$^2$ COSMOS area in the sky. We inspected the COSMOS2015 deep 
catalogs in search for galaxies close to our object. We found two candidates in 
a 5\farcs0 radius circle around the object position. These galaxies, designated 
as A and B, at distances of 1\farcs5 South and 2\farcs5 North-West respectively 
from the SN location, are shown in Figure \ref{fig:hosts}.
Galaxy B, a clearly bright and extended source, is detected in the NICMOS F160W 
image and was used as a model to determine the upper limit of the host.

The photometry of galaxy B was already measured in previous COSMOS surveys:
\citet{capak07, ilbert09}, yielding a photometric redshift estimate of
$z=0.53$. The photometric redshift of A is poorly constrained, having a median
of $z=1.61$ and a peak of probability at $z=0.37$. COSMOS2015 reports the
median as the \emph{photo-z} result calculating a 68\% confidence interval
around it. Similarly, other fitted quantities such as stellar mass,
star formation rate and age are given for both values, but the error bars are
only reported for those coming from the median of the $z$ probability
distribution.

In Section \ref{sec:lc_interpolation} we show that it is unlikely for UID
30901 to have a redshift larger than one. In such a case, we would be observing 
the near and far UV rest-frame light curves in the $g$ and $r$ filters 
respectively, but the radiation measured in those bands does not resemble the 
expected abrupt decrease at those wavelengths for this (or any) kind of SN.
Similarly, from our blackbody fitting, we find that photospheric temperatures 
consistent with such a redshift would be large (between 20,000 and 30,000 K) 
during the first couple of weeks of observations. Besides, $z=1.0$ would imply 
an absolute magnitude for UID 30901 $M_{\rm abs} \lesssim -23.5$ mag, while no 
SN event of such luminosity has been ever observed. Hence, if galaxy A is found
at the same redshift as the host of UID 30901, we can reject a $z=1.61$ value. Instead, we adopt the peak redshift of $z=0.37$ as representative.



Having tentative redshifts for the SLSN of 0.53 and 0.37, and an upper limit of $z < 1$, we tested whether the F814W magnitude and the F160W upper limits are consistent with these $z$ values. Using the Flexible Stellar Population Synthesis code \citep{conroy09,conroy10}, we obtained synthetic magnitudes for the host of UID 30901 for a redshift range of $0.1 \le z \le 0.9$ after assuming the usual parameters for SLSN hosts as described by \cite{perley16}. That is, an old population that contributes to a stellar mass in the $10^{6-8}$ M$_{\odot}$ range, a recent burst with constant star formation rate (SFR) of $0.03 - 3$ M$_{\odot} $yr$^{-1}$, ages for the burst of $10 - 500$ Myr, and a modest amount of extinction of A$_V$ = 0.0, 0.1 and 0.5 magnitudes.

We find that given the faint magnitudes of our host, only small host masses ($10^{6}$ M$_{\odot}$), low SFRs ($0.03$ $_{\odot}$yr$^{-1}$) and redshifts below 0.5 are in agreement with our data. The burst duration is the variable that introduces most of the vertical scatter to the trend seen in Figure \ref{fig:hosts}, followed by extinction. We also find that the point-like upper limit for the F160W filter is found in agreement with the synthetic data, implying that the host of UID 30901 could be a very compact source.

Our previous results agree well with the redshifts found for galaxies A and B, which have projected distances of 7 and 15 kpc to the SLSN host, respectively, for assumed redshifts of 0.37 and 0.53. For the remainder of this work, we will adopt these redshifts as representative candidate-$z$ values for the host of our SLSN. Coordinates and photometry of the host and possible companion galaxies are detailed in Table \ref{tab:hosts}.

\begin{table}
    \centering
        \begin{tabular}{|l|c|c|c|}
        \hline
        \hline
        \multicolumn{3}{c}{Host Galaxy}\\
        \hline
        Parameters & & \\
        \hline
        RA & \multicolumn{2}{c}{149.821917}\\
        DEC & \multicolumn{2}{c}{2.060639}\\
        mag F814W & \multicolumn{2}{c}{27.30(17)}\\ 
        mag F160W & \multicolumn{2}{c}{.......}\\
        \hline
        \hline
        
        \multicolumn{3}{c}{Companion Galaxies}\\
        \hline
        Parameters   & A & B\\             
        \hline
        RA & 149.821884 & 149.821243 \\
        DEC & 2.06029 & 2.060975\\
        Ang. Dist. & 1.45" & 2.54"\\
        mag $g$ & 26.98(28) & 25.70(10) \\
        mag $r$&$\dotfill$ &24.90(06) \\
        mag $i$ & 26.96(18) & 24.56(04) \\
        mag $z$ & 26.20(20) & 24.41(07) \\
        mag $y$& $\dotfill$    & 24.24(13) \\
       
       \hline
        \multicolumn{3}{c}{UltraVISTA}\\
        mag $Y$  & 26.09(12)& 24.27(03)\\
        mag $J$  & 25.45(10)& 24.29(05)\\
        mag $H$  & 25.59(18)& 24.57(10)\\
        mag $K$  & 25.41(15)& 24.34(08)\\
        \hline
        \multicolumn{3}{c}{SuprimeCam}\\
        mag $B$  & 27.65(23)& 26.31(10)\\
        mag $V$  & 27.39(30)& 25.59(09)\\
        mag $r$  & 26.93(19)& 24.98(05)\\
        mag $i$  & 27.19(33)& 24.67(05)\\
        mag $z$  & 26.42(18)& 24.44(04)\\
        \hline
        \multicolumn{3}{c}{COSMOS2015$\dag$}\\
        ID$\dag$ & 499317 & 500106 \\
        photo-z median & 1.61$_{-0.52}^{+0.69}$& 0.52(04) \\
        photo-z best-fit & 0.37& 0.53 \\
        Age [years]& $5\times 10^7$& $9.05\times 10^8$\\
        SFR [log$_{10}$] & 0.59& -1.29\\ 
        sSFR [log$_{10}$] & -7.34 & -9.57\\
        Proj. Dist [Kpc]&7.4&15.4 \\
        Stellar Mass [log$_{10}$ $M_{\odot}$] & 7.93 & 8.28 \\
        
        \hline
        \end{tabular}
    \begin{tablenotes}
       \item[*] $\dag$: \cite{laigle16}. Only best-fit
       derived quantities are considered (see text).
      \end{tablenotes}
    \caption{Properties of the host and possible companion galaxies. Top: we present coordinates and photometry measured from the single image of HST. 
    Middle: we present coordinates, angular distance to the SN position, and photometry we measured from SUBARU Hyper SuprimeCam optical images taken when
    the SN was not present. Bottom: COSMOS2015 photometry and resulting parameters from 
    SED fittings. The photometry includes deep imaging from UltraVISTA and SUBARU SuprimeCam among other sources.}
    \label{tab:hosts}
\end{table}

\subsection{Explosion date: Power Law and Linear Fits} 
\label{sec:exp_time}
\subsubsection{Rise Time Fits}

A power law model for the flux $F \propto (t-t_{0})^{\alpha}$
has been used often to characterise the early part of a diverse
number of transients. In our case, the power law provides a good fit to
the rising part of the NIR data, and therefore can be used to estimate
the explosion date.

To fit a power law we first transformed our $YJHK_{s}$-band
photometry to flux units. Next, we fit the early $YJHK_{s}$-band 
light curves simultaneously with a six parameter power law model: 
four constants (one for each band) the time of explosion ($t_{0}$) 
and the power law exponent ($\alpha$). We also fit a linear 
model to the flux, due to the linear behaviour seen
in the middle section of the rise time. In both cases we used a MCMC code
based on the \textit{emcee} python package \citep{foreman13}. The
results for both fits are summarised in Table \ref{tab:plaw} and
a plot of the fit is shown in Figure \ref{fig:plaw}.

Our power law model places the explosion date $\sim$ 6 days before the
first detection, while the linear fit places the explosion $\sim$ 15 
days earlier. Both estimates are in the observer frame. 
We did not include the optical photometry in the fitting as it is not well 
sampled in the early phase of the light curves. In what follows we 
will compare these extrapolations with the other observational constraints.

\subsubsection{Explosion Date}

\begin{figure}
\includegraphics[width=8cm]{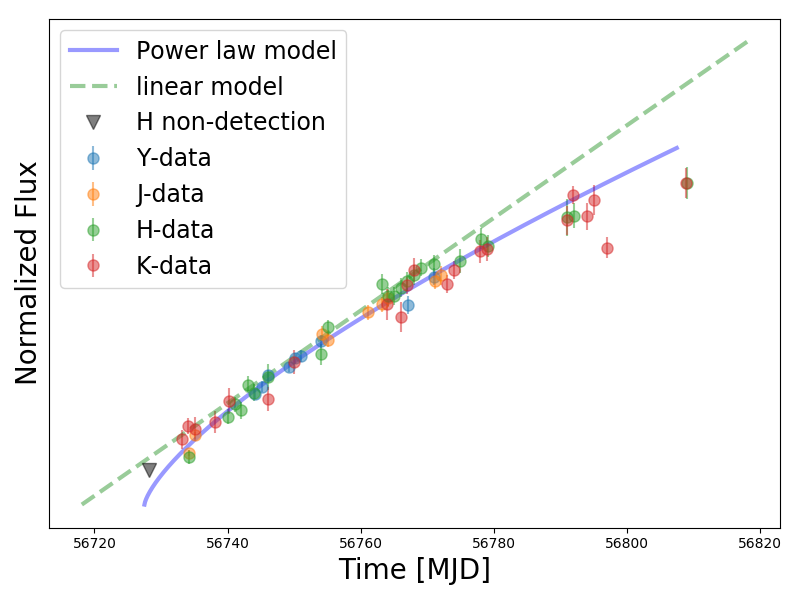}
\caption{Linear and Power Law fit of the early part of the light curve.}
\label{fig:plaw}
\end{figure}

UID 30901 was first detected on March 17th, 2014 at 03:10 UT time 
in the $H$ band by UltraVISTA using VIRCAM on the VISTA telescope at
Paranal and 10 minutes later was detected by DECam in $g$ band mounted
on Blanco telescope at CTIO. A deep VISTA
$H$-band image taken on March 12th at 03:21 UT time
shows no signal of the SN. We added artificial stars to the 
non-detection image of diverse brightness and tried to detect them 
using the DAOFIND algorithm \citep{stetson87}, looking for the brightness
value that reaches a 50\% of detection rate. With this process, we estimated 
a limiting magnitude of 23.3 in the deep $H$-band image (see Figure \ref{fig:plaw}).
This corresponds to less than 10\% of the $H$-band flux at 
peak and is $\sim 0.1$ mag below the linear fit to the early
part of the light curve. 

Based on our fits of the early data points and the limiting magnitude 
obtained five days before the first detection, we estimate an 
explosion epoch of March 10th, 2014.

\begin{table}
    \caption{Early part of the light curves.}
    \begin{center}
    \begin{tabular}{|l|c|}
    \hline
    \multicolumn{2}{|c|}{Linear Fit}\\
    \hline
    $t_0$ [MJD]& 56718.5(6)\\
    \hline
    \multicolumn{2}{|c|}{Power Law}\\
    \hline
    $t_0$ [MJD] & 56727.5(5)\\
    $\alpha$ & 0.72(03)\\
    \hline
    \multicolumn{2}{|c|}{Last Non-Detection}\\
    \hline
    $t_{ND}$ [MJD] & 56728.14 \\
    \hline
    \multicolumn{2}{|c|}{First Detection}\\
    \hline
    $t_1$ [MJD] & 56733.13\\
    \hline
    \end{tabular}
    \end{center}
    \label{tab:plaw}
\end{table}

\subsection{Comparison with other supernovae}

To get further insights into the nature of UID 30901, we compare its $r$-band
absolute magnitude light curve with low-z stripped-envelope SNe from the
Carnegie Supernova Project I \citep{hamuy06, stritzinger18a}, with
SLSNe from the Pan-STARRS1 Medium Deep Survey \citep{lunnan18} and
the Dark Energy Survey \citep{angus19}. We also compare the UID 30901 spectral
energy distribution (SED) with the spectra of superluminous SNe at two
representative epochs separated by more than a hundred days: near maximum light
and at a later phase.

\subsubsection{Light curve comparison}

Looking carefully at the light curve evolution of SN UID\,30901 (see Figure
\ref{fig:opnir_lc}) we can readily dismiss the hypothesis that it could
correspond to a SN Ia or to a normal SN\,II. This is based on the fact that
SNe Ia peak earlier in the NIR than in the optical \citep[see][]{folatelli10},
they have narrower light curves than UID\,30901, with typical rise times of
about 15 to 20 days in the rest-frame \citep[see e.g.,][]{firth15}, and
display distinctive double peaked light curves in the NIR. None of these
characteristics are observed in the light curves of UID\,30901. Similarly,
normal type II SNe display a rise time to maximum in the optical of less than
30\,days, and they often display a `plateau' phase followed by a rapid decline
once the hydrogen-rich ejecta has recombined to finally settle in a radioactive
tail at about 80 to 150\,days after the explosion \citep[see e.g.,][for more
  details]{anderson14}. Again, these characteristics are not observed in the
UID\,30901 light curves (see Figure \ref{fig:opnir_lc}).

\begin{figure}
    \includegraphics[width=8cm]{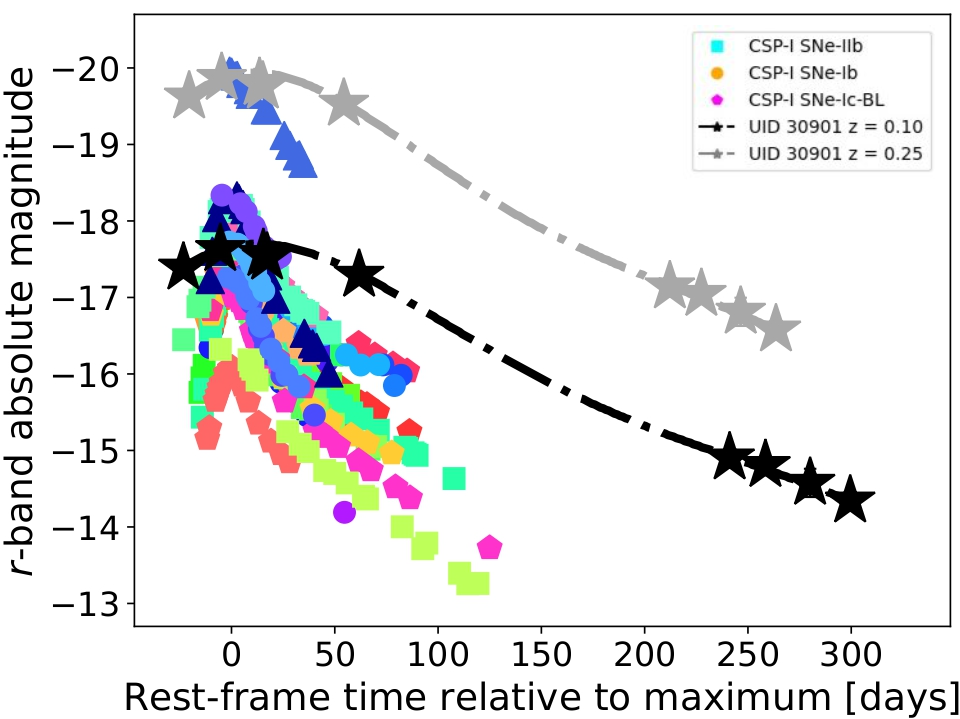}
    \caption{Comparison between the $r$-band absolute magnitude of UID\,30901
      and the stripped-envelope SNe from CSP-I \citep{stritzinger18a,taddia18}
      after assuming redshifts of $z=0.10$ (black stars) and $z=0.25$ (grey
      stars), to make its luminosity at maximum consistent with a normal SN Ic
      and a relatively bright Ic-BL, respectively. In the figure SNe type IIb
      are shown as squares, type Ib as pentagons, type Ic as circles and type
      Ic-BL as triangles. As can be seen the light curve of UID\,30901 is
      broader than any of these SN types.}
        \label{fig:SE_lc_comparison}
\end{figure}

The comparison with Stripped Envelope SNe can be seen more clearly in Figure
\ref{fig:SE_lc_comparison}, where we plot the $r$-band light curve of
UID\,30901 and the optical photometry of stripped-envelope SNe from the CSP-I
sample \citep{stritzinger18a,taddia18}. The redshift value required to make
the $r$-band maximum brightness of UID\,30901 consistent with the mean
$r$-band absolute magnitude of the Type Ic CSP-I subsample ($M_{r} = -17.7\pm
0.2$ mag) is $z \approx 0.1$. Similarly, $z \approx 0.25$ makes the
$r$-band maximum brightness of UID\,30901 consistent with a bright SN Ic-BL
($M_{r} \approx -19.9$ mag). All the relevant parameters such as the time of
$r$-band maximum, redshift, luminosity distance, and Galactic and host galaxy
extinction values for the CSP-I SNe were adopted following \citet{taddia18}.
The host galaxy reddening reported by \citet{taddia18} was computed following
the methodology presented by \citet{stritzinger18b}. 
As can be seen, the light curve of UID\,30901 is
significantly broader than any of these SN types, rejecting the possibility of
a normal luminosity stripped-envelope SN origin for UID\,30901, and suggesting a
longer diffusion timescale and therefore a more massive SN ejecta for this
object.

\begin{figure}
    \includegraphics[width=8cm]{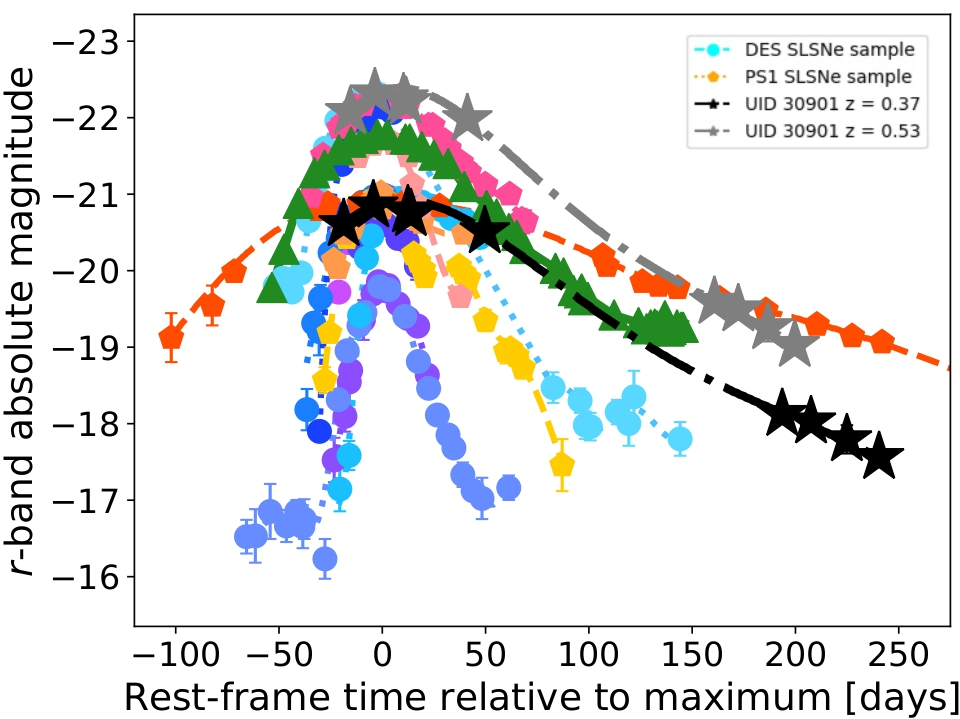}
    \caption{Comparison between the $r$-band absolute magnitude of UID\,30901
      and SLSNe from PS1 \citep{lunnan18}, DES \citep{angus19} and SN\,2006gy
      \citep{smith07}. For PS1 and DES SLSNe we show the Gaussian process
      $r$-band light curve interpolations presented in \citet{cartier22}. PS1
      SLSNe are shown as pentagons and their light curve interpolations with
      dotted lines, DES SLSNe are shown as circles and their light curve
      interpolations using dashed lines, SN\,2006gy is shown as green
      triangles and the host galaxy distance and reddening are from
      \citep{smith07}. We show SN UID\,30901 assuming at z=0.37 (black stars)
      and at z=0.53 (grey stars)}.
    \label{fig:SL_lc_comparison}
\end{figure}

Next, we compare the light curve of UID\,30901 with other SLSNe to confirm its
superluminous nature. In Figure \ref{fig:SL_lc_comparison}, the $r$-band
absolute magnitude light curve of UID\,30901 is presented when assuming
redshifts of $z=0.37$ and $z=0.53$ (see Section \ref{sec:host}). These are compared
with SLSNe from PS1 \citep{lunnan18} and DES \citep{angus19} for the redshift
range $0.25 \leq z \leq 0.65$ to avoid the need of k-corrections, and the Type
II superluminous SN\,2006gy \citep{smith07}. The time of maximum light for the
PS1 and DES SNe was computed using Gaussian process interpolation as 
described in \citet{cartier22}, and the light curves were corrected by
Galactic reddening ($E(B-V)$) as reported by \citet{Schlafly11} using the
\citet{Cardelli89} extinction law. For SN\,2006gy the $A_{R}$ values reported
by \citet{smith07} were used for the reddening correction, this is $A_{R}=
0.43$ and $A_{R}=1.25$ for our galaxy and the SN host galaxy, respectively. No k-corrections were applied to the light curves as we expect this correction to be of the order of a few tenths of magnitude for the highest redshift objects, which is small compared to the dispersion observed in the maximum luminosity and with the light curve shape diversity displayed by SLSNe
\citep[see][]{decia18,lunnan18,angus19,cartier22}.


Looking at Figure \ref{fig:SL_lc_comparison} we find that the shapes of
luminous ($M_{r} < -20$\,mag) SLSNe are similar to UID\,30901, thus
confirming its superluminous nature. UID\,30901 is among the objects with the
broadest light curves, regardless of the redshift assumed for this SN. Aside
from a small group of fainter SLSNe with maximum brightness $M_{r} \gtrsim
-20$\,mag, showing light curves shapes that could be consistent with a bright SN
Ic-BL, all SLSNe with broad light curves such as UID\,30901 have maximum
luminosities $M_{r} < -20$\,mag. Unless UID\,30901 is a very peculiar
object with a very broad light curve and a moderate luminosity, we can safely
assume that UID\,30901 has a maximum absolute magnitude $M_{r} <
-20$\,mag, implying that its redshift must be $z > 0.25$.

A review of the literature reveals that the brightest SLSNe reported have
absolute magnitudes in the range of $-22.0$\,mag\,$\lessapprox
M_{\mathrm{abs}} \lessapprox -22.5$\,mag
\citep{smith07,vreeswijk14,smith16,decia18,smith18,lunnan18,cartier22,yin21}.
Therefore assuming that UID\,30901 reached one of the brightest maximum
luminosities reported for a SLSNe ($M_{r} \approx -22.5$), we can place an
approximate redshift upper limit of $z \lesssim 0.7$ for this SN.


\begin{figure*}
    \includegraphics[width=\textwidth]{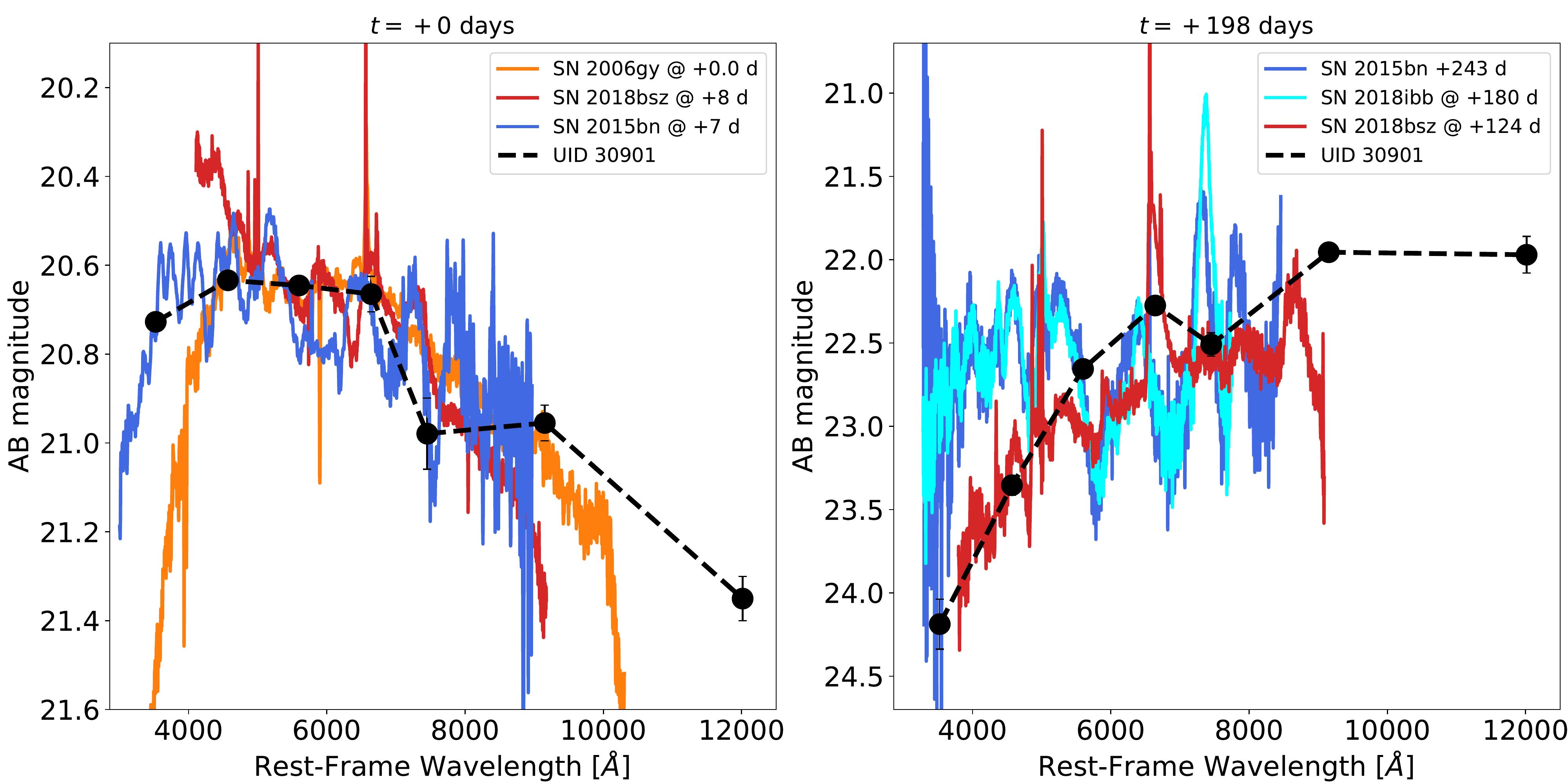}
    \caption{SED comparison between UID\,30901 and other SLSNe assuming
      z=0.37. In the left-panel the SED of UID\,30901, the spectra of SLSNe-I
      SN\,2015bn \citep{nicholl2016} and SN\,2018bsz \citep{Anderson18bsz},
      and the spectrum of the SLSN-II SN\,2006gy \citep{smith07} are compared
      within days of maximum light. In the right-panel the late phase SED of
      UID\,30901 (+198\,days) and the late time spectra of SLSNe SN\,2015bn
      \citep{nicholl2016}, SN\,2018ibb, SN\,2018bsz are compared. It can be seen that
      after maximum light SN\,2018bsz starts to interact with a dense CSM
      displaying a strong H$\alpha$ emission line and evidence for dust
      condensation \citep[see][]{Chen22}. The late spectra of SN\,2018bsz and
      SN\,2018ibb were obtained with the Goodman spectrograph mounted at SOAR
      telescope, and are presented here for comparison but will be published
      elsewhere.}
    \label{fig:SED_comparison_zp37}
\end{figure*}

\begin{figure*}
    \includegraphics[width=\textwidth]{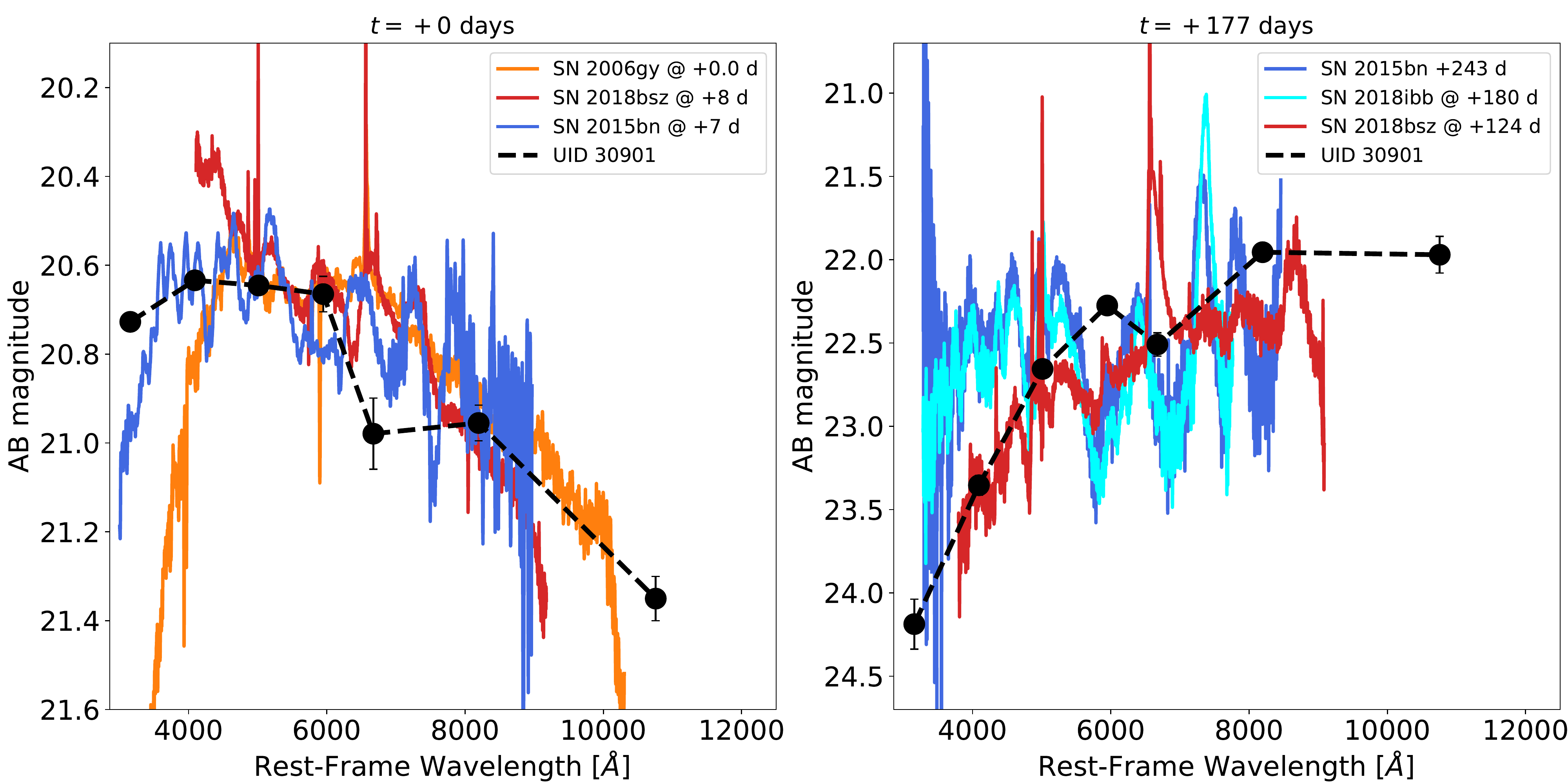}
    \caption{Same as Figure \ref{fig:SED_comparison_zp37}, but in this case assuming z=0.53.} 
    \label{fig:SED_comparison_zp53}
\end{figure*}

\subsubsection{Spectral energy distribution }
\label{sec:3.3.2}
The comparison of the light curve of UID\,30901 with the diversity of SN types 
has shown that this object displays a broad light curve, which is only 
consistent with a SLSN. From this, we have placed constraints on the redshift 
of this object to be in the range of $0.25 < z \lesssim 0.7$. This
redshift range is in agreement with the photometric redshifts of the potential
host galaxies discussed in Section \ref{sec:host}. In the following, we will
compare the $grizYJH$ SED of UID\,30901, at $\mathrm{MJD}=56764$ (at about
maximum bolometric luminosity) and at $\mathrm{MJD}=57035$ (+271\,d after
maximum in the observer frame), with other SLSNs using the redshift values for
the potential host galaxies from Section \ref{sec:host}.

In the left panel of Figure \ref{fig:SED_comparison_zp37} we compare the SED
of UID\,30901 after assuming $z=0.37$ with the spectra of SLSNe SN\,2015bn
\citep{nicholl2016}, SN\,2018bsz \citep{Anderson18bsz} and SN\,2006gy near the
time of maximum luminosity. At this phase the SED of UID\,30901 is well
reproduced by a blackbody (see Section \ref{sec:bol_lumin}) and it is also in
very good agreement with the spectrum of SN\,2015bn, while the near maximum
spectra of SN\,2018bsz \citep{Anderson18bsz} appears too blue, and the spectra
of SN\,2006gy seems too red, even after correcting them by Galactic and host
galaxy reddening. Assuming $z\simeq 0.37$, the $Y$-band flux drop in
UID\,30901 coincides with the strong \ion{O}{i} $\lambda 7774$ absorption line
in the spectrum of SN\,2015bn. In Figure \ref{fig:SED_comparison_zp53} a similar comparison is presented but assuming $z=0.53$, where the SED of
UID\,30901 is again similar to SN\,2015bn, but at this redshift the drop in
the $Y$ band no longer coincides with the \ion{O}{i} $\lambda 7774$ absorption
line. The overall spectrum of SN\,2015bn coincides better with UID\,30901
assuming $z=0.37$.

In the right panel of Figure \ref{fig:SED_comparison_zp37} the SED of
UID\,30901 is compared with the spectra of SN\,2015bn \citep{nicholl16b}, SN\,2018bsz and SN\,2018ibb. The spectra of SN\,2018bsz and SN\,2018ibb were obtained with the Goodman spectrograph mounted at the SOAR telescope, and are presented here for comparison but will be published elsewhere. Assuming $z=0.37$, the SED of UID\,30901 shown in the right panel of Figure \ref{fig:SED_comparison_zp37} corresponding to $+200$\,d in the rest frame, and it has evolved dramatically becoming very red below 6500 \AA\ when compared to the SED at maximum light. The spectra of SN\,2015bn or SN\,2018ibb, which are representative of a ``{\it normal}" SLSN-I at a late phase no longer provide the best comparison to the SED of UID\,30901 at this phase. These two SNe are bluer by nearly one mag at 4000 \AA\ compared with UID\,30901. On the other hand, the spectrum of SN\,2018bsz at +124\, days provides a good comparison to the SED of UID\,30901. SN\,2018bsz was classified as the closest SLSN I, at a redshift of $z=0.0267$ \citep[see][]{Anderson18bsz}. It showed some unusual features including a long plateau before maximum light. After maximum, it started to show a strong and broad H$\alpha$ emission, evidence of ejecta-CSM interaction, and also indications of dust formation \citep{yan17}.  In the right panel of Figure 
\ref{fig:SED_comparison_zp37}, the $z$-band of UID\,30901 coincides and provides a good match to the H$\alpha$ emission in SN\,2018bsz. Assuming $z=0.53$ (right panel of Figure \ref{fig:SED_comparison_zp53}), SN\,2018bsz also provides the best comparison to the UID\,30901 SED, but the agreement is not as good as assuming $z=0.37$.

\subsection{Characterising the light curves}
\label{sec:lc_interpolation}

To characterise the light curves of UID 30901 we used a simple polynomial
fitting to interpolate the photometry. Three phases were distinguished in our
light curves. The early phase comprises the rise to maximum, the time of the
maximum, and in some optical bands the beginning of a decline from the peak
brightness. During the second phase, between +200 and +380 days post peak,
the SN shows a nearly linear decline, as we will show below. Finally, the last
phase comprises observations from +594 days to our last detection at +685 days 
observer frame, where the SN seems to decline at a slower rate compared to the 
previous phase.

\subsubsection{Early Phase}
\begin{figure}
    \centering
    \includegraphics[width=8cm]{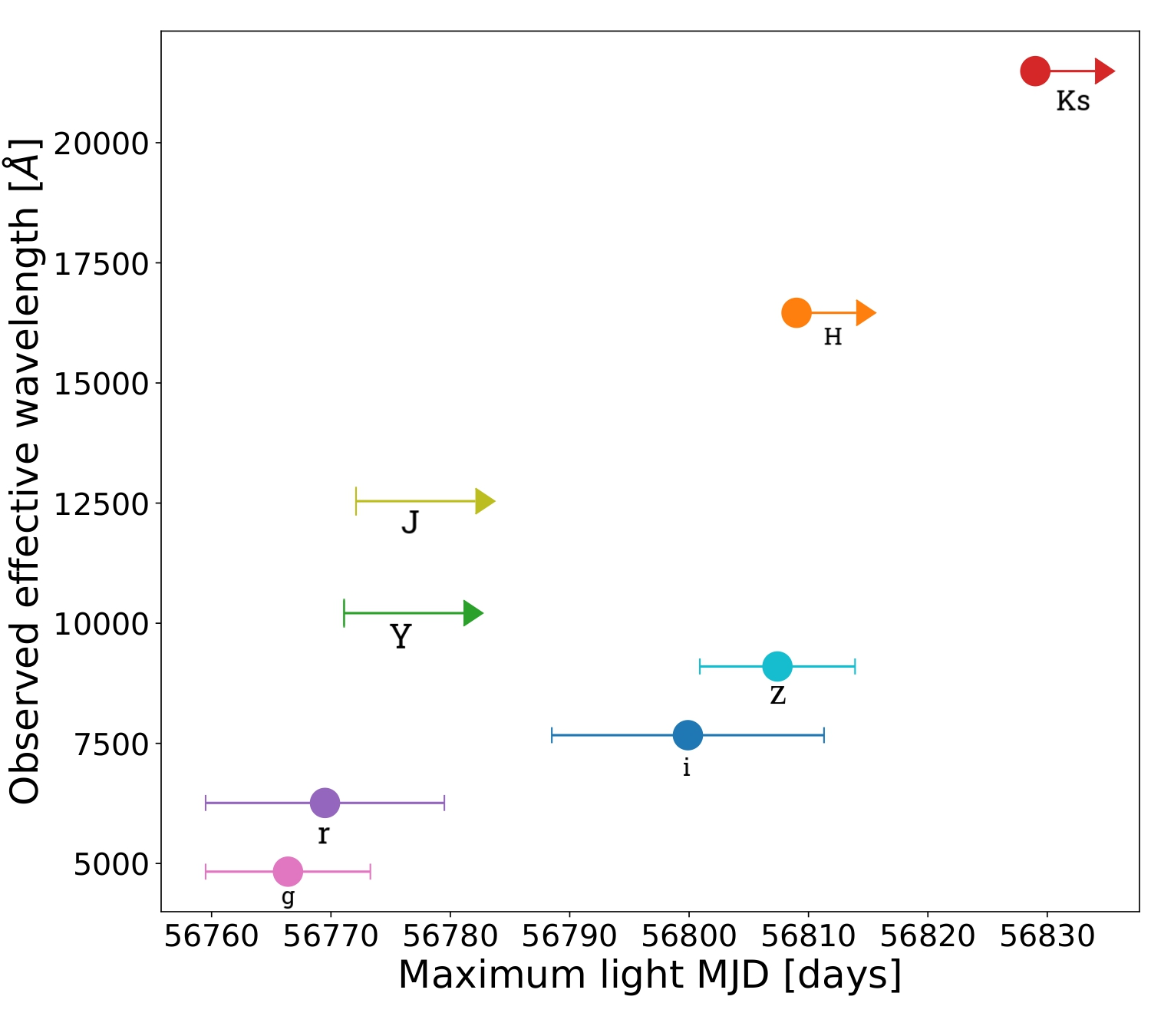}
    \caption{Epoch of maximum light for different bandpasses. For the NIR bands 
    we present a lower limit of the time of maximum light, observations of
    $H$ (in orange) and $K_{s}$ (in red) bands seem to be very close
    to peak. Our estimates for the epoch maximum light suggests an
    increase of the time of maximum with wavelength.}
    \label{fig:time_of_max}
\end{figure}

To characterise the rise to the peak and the decline of the light
curves of UID 30901, we fit a low order polynomial. These fits
provide an estimated epoch of the maximum and the peak magnitude
in the optical bands. The NIR light curves do not
have any photometric points beyond maximum and hence no early
decline is observed. However, in the $H$ and $K_{s}$ bands the
light curves seem to reach close to the peak. Hence, for the 
NIR bands we can place lower limits in MJD and upper limits in 
magnitudes for the maximum. We report our results 
in Table \ref{tab:peak_mag} and in Figure \ref{fig:time_of_max}.

As can be seen in Figure \ref{fig:time_of_max} our multi-wavelength coverage suggests a relation between the epoch of peak brightness and the observed effective wavelength, where longer effective wavelengths peak later. This relation seems to extend to the NIR bands, where the NIR bands reach their peak brightness after the optical bands.  We note that this behaviour of reaching maximum brightness later at longer effective wavelengths is similar to the one reported in well-observed SNe Ic and Ic-BL \citep[e.g.,][]{hunter09, pignata11, taddia18} and as noted in the previous section is different from the behaviour observed in SNe Ia, where the NIR bands ($iYJHK_{s}$) reach their peak brightness between 3 to 5 days before the $B$-band maximum \citep[which is similar to the time of the $g$-band maximum;
  see][]{folatelli10}.

\begin{table}{}
    \caption{Peak light curve information for UID 30901.}
    \label{tab:peak_mag}
    \begin{tabular}{@{}lcccc}
    \hline
    Filter & MJD & Observed & z=0.37 & z=0.53\\ 
           & (days) & (mag) & (mag) & (mag)\\
    \hline \hline
    $g$    & $56766$($07$)  & $20.78$($05$) & -20.64 & -21.63 \\
    $r$    & $56770$($10$)  & $20.67$($04$) & -20.75 & -21.66 \\
    $i$    & $56800$($11$)  & $20.36$($13$) & -21.06 & -22.05 \\
    $z$    & $56807$($07$)  & $20.34$($07$) & -21.08 & -22.07 \\
    $Y$    & $> 56771.1$       & $< 20.77$       & $<-20.65$ & $<-21.64$ \\
    $J$    & $> 56772.1$       & $< 20.86$       & $<-20.56$ & $<-21.55$ \\
    $H$    & $\geq 56809.0$    & $\leq 20.89$    & $\leq -20.53$ & $\leq -21.52$ \\
    $K_{s}$& $\geq 56829.0$    & $\leq 21.05$    & $\leq -20.37$ & $\leq -21.36$ \\
    
    \hline
    \end{tabular}
    \begin{tablenotes}
    \item Numbers in parenthesis correspond to 1-$\sigma$ statistical uncertainties. The given peak absolute magnitudes correspond 
    just to apparent magnitude minus the distance modulus in a $\Lambda CDM$
    Universe with $\Omega_M=0.3$, $\Omega_{\Lambda}=0.7$ and Hubble-Lemaitre parameter $h=0.7$.
\end{tablenotes}
\end{table}

\subsubsection{Linear Decline Phase} \label{sec:linear_interpol}

From November 2014 to May 2015 the light curves of UID 30901 display a post
maximum linear decline, hence we fit a linear model to measure the decline
rate, interpolate, and in some cases extrapolate the light curves over this
period. To assess the goodness-of-fit of the linear model we computed
the reduced chi-squared ($\chi^{2}_{\nu}$) and the root mean squared
($\mathrm{\it rms}$).

\begin{figure}
\includegraphics[width=8cm]{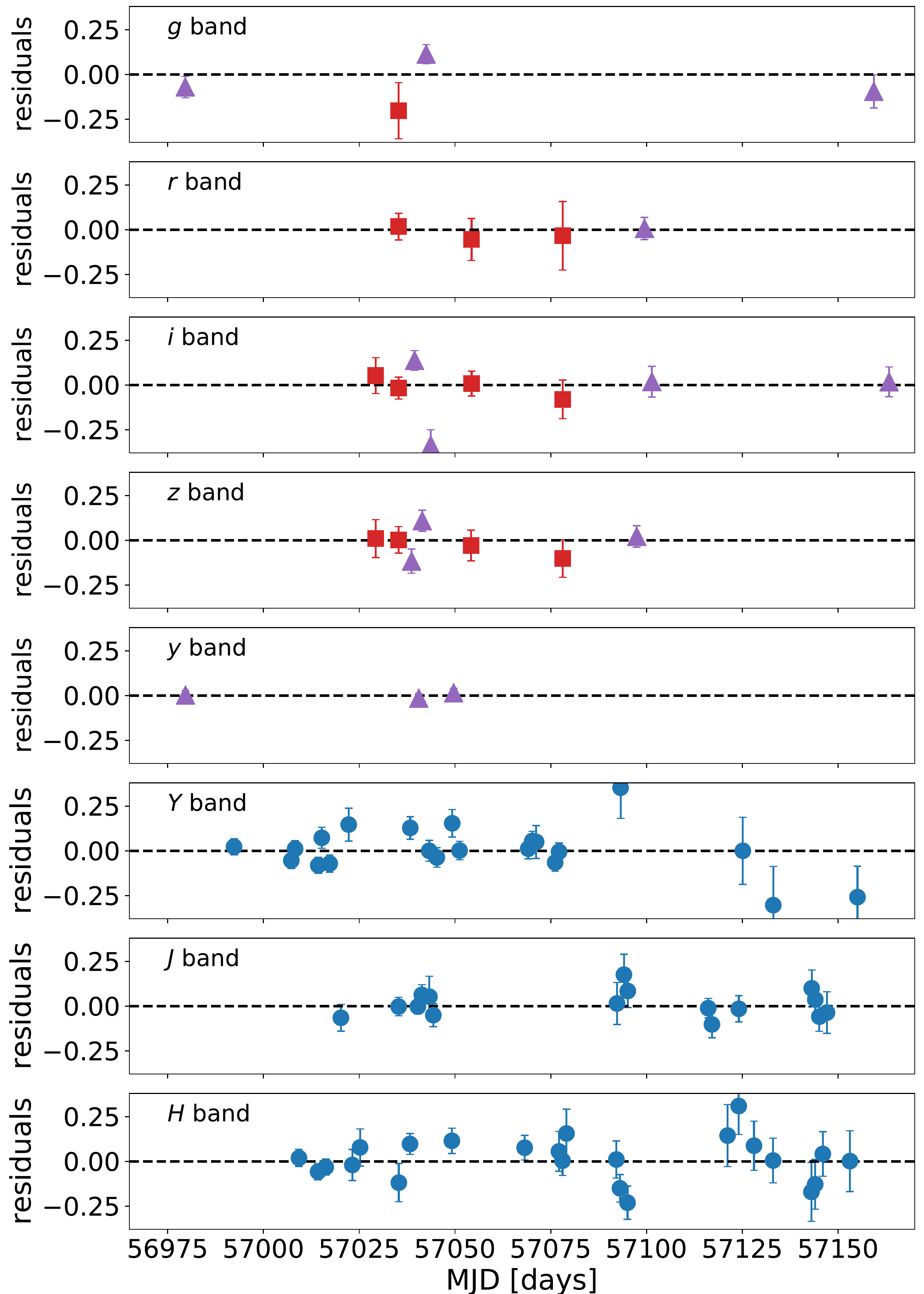}
\caption{Residuals from the linear fits to the light curves of UID 30901
from November 2014 to May 2015. We present the DECam/Blanco ($griz$) 
residuals as red squares, the ($grizy$) HSC/SUBARU residuals as purple 
triangles, and the VISTA ($YJH$) residuals as blue circles. In the 
top-left of each panel we indicate the corresponding band.}
\label{fig:tail_residuals}
\end{figure}

We summarise the decline rates of the $grizyYJH$ bands, the $\chi^{2}_{\nu}$
and the $\mathrm{\it rms}$ of the fits in Table \ref{tab:linear_decline}, and
we present the linear model residuals in Figure \ref{fig:tail_residuals}.
Although a linear fit may be considered a too simplistic model, the residuals,
the $\chi^{2}_{\nu}$, and the $\mathrm{\it rms}$ indicate that a linear model
is an adequate approximation to the SN luminosity decline over this phase (see
Figure \ref{fig:tail_residuals}).



\begin{table*}
\begin{center}
\caption{Linear decline rate information for UID 30901.}
\label{tab:linear_decline}
\begin{tabular}{@{}lccccccc}
\hline
Filter & Observed decline rate & Decline rate at $z=0.37$ & Decline rate at $z=0.53$ & $\chi^{2}_{\nu}$ & $\mathrm{\it rms}$  & $n_{\mathrm{obs}}$ \\
       & (mag/100\,days)       & (mag/100\,days)          & (mag/100\,days)          &                  & (mag)           &                    \\
\hline
\hline
$g$    & $1.09$($0.06$) & $1.49$($0.09$) & $1.67$($0.09$) & $4.26$ & $0.087$ & $4$  \\
$r$    & $0.89$($0.15$) & $1.22$($0.21$) & $1.37$($0.23$) & $0.16$ & $0.022$ & $4$  \\
$i$    & $0.72$($0.07$) & $0.98$($0.09$) & $1.10$($0.10$) & $3.76$ & $0.054$ & $8$  \\
$z$    & $0.89$($0.11$) & $1.17$($0.15$) & $1.31$($0.17$) & $1.52$ & $0.032$ & $7$  \\
$y$    & $0.90$($0.05$) & $1.24$($0.06$) & $1.38$($0.07$) & $0.76$ & $0.011$ & $3$  \\
$Y$    & $0.97$($0.04$) & $1.33$($0.06$) & $1.48$($0.06$) & $1.66$ & $0.031$ & $21$ \\
$J$    & $0.72$($0.04$) & $0.98$($0.06$) & $1.10$($0.06$) & $0.70$ & $0.019$ & $16$ \\
$H$    & $0.63$($0.04$) & $0.87$($0.06$) & $0.97$($0.06$) & $1.40$ & $0.026$ & $23$ \\
\hline
\end{tabular}

\begin{tablenotes}
\item Numbers in parenthesis correspond to 1-$\sigma$ statistical uncertainties.
\end{tablenotes}
\end{center}
\end{table*}

Recently \citet{decia18} analysed a sample of 26 spectroscopically confirmed
hydrogen-poor SLSNe from the (i)PTF survey. They measured the rest-frame
$g$-band decline rate after 60 days from the $g$-band maximum and found a
range of decline rates from $\sim 0.25$ to $\sim 2.25$ mag/100\,days for their
sample, with a mean of $1.3 \pm 0.5$ mag/100\,days. If UID 30901 is at $z
\simeq 0.37$, the $r$-band would be similar to the rest-frame $g$-band, and
the measured $r$-band decline rate would be $1.22 \pm 0.21$ mag/100\,days
agreeing very well with the mean decline rate of the SLSN sample analysed by
\citet{decia18}. If UID 30901 is at $z \simeq 0.53$, the rest-frame $g$-band
would cover part of the $r$ and $i$ bands, with the rest-frame effective
wavelength found in the $i$-band. The decline rate values are found to be
$1.37 \pm 0.23$ and $1.10 \pm 0.10$ for the $r$ and $i$ bands, respectively,
which compare well with the rest-frame $g$-band decline rates reported by
\citet{decia18}.

\subsubsection{Final Year Characterisation and Flattening at Late Epochs}	
\label{sec:FinalYearandFlattening}

We have sparse observations from December 2015 to March 2016 in 
several bands. Some of them are the results of combining several
frames over about a month to obtain a SN detection at very faint
magnitudes, as is the case of the $iyJ$-bands. 

\begin{table}
\begin{center}
\caption{Linear decline rate information for the final year of UID 30901.}
\label{tab:late_decline}
\begin{tabular}{@{}lcccc}
\hline
Band & Observed & $z=0.37$ & $z=0.53$ & Brightness at \\
     & (mag/100 days) & (mag/100 days) & (mag/100 days) & MJD=57428.0 \\
     & & & & (mag) \\
\hline \hline
$z$    & $0.5$($0.3$) & $0.7$($0.4$) & $0.8$($0.4$) & $25.0$($0.1$) \\
$J$    & $0.1$($0.2$) & $0.2$($0.3$) & $0.2$($0.3$) & $24.4$($0.1$) \\
\hline
\end{tabular}
\begin{tablenotes}
\item Numbers in parenthesis correspond to 1-$\sigma$ statistical
  uncertainties. Two and four epochs were used for the $z$ and $J$ bands, respectively
\end{tablenotes}
\end{center}
\end{table}

\begin{table}
\centering
\caption{Difference between observed magnitudes in the last observing season and a linear fit extended from the linear decay phase.}
    \label{tab:mag_diff}
\begin{tabular}{@{}lcccccc}
\hline
Band   &  MJD	& Residuals	& $\sigma_{phot}$ & $\sigma_{fit}$ & $\sigma_{magtot}$ & $\sigma$ \\
       &  	&(mag)&(mag)&(mag)&(mag)&(mag) \\
\hline
\hline
$g$      &  57454.329	&          -2.0		& 0.18 &	   0.26 &	     0.31 &      6.6 \\
$i$      &  57428.0	&          -0.5		& 0.19 &	   0.25 &            0.31 &      1.7 \\
$z$      &  57402.669   &          -0.6		& 0.12 &	   0.39 &	     0.41 &      1.5 \\
$z$      &  57459.474	&          -0.8		& 0.12 &	   0.45 &	     0.47 &      1.7 \\
$y$      &  57440.0	&          -0.4		& 0.21 &	   0.19 &	     0.28 &      1.5 \\
$J$      &  57374.099	&          -0.0		& 0.13 &	   0.12 &	     0.18 &      0.04 \\
$J$      &  57408.758	&          -0.4		& 0.09 &	   0.13 &	     0.16 &      2.3 \\
$J$      &  57425.337	&          -0.4		& 0.16 &	   0.16 &	     0.22 &      1.6 \\
$J$      &  57457.943	&          -0.5		& 0.14 &	   0.18 &	     0.23 &      2.3 \\
\hline
\end{tabular}
\end{table}

\begin{table}
\centering
\caption{Decline rate at second and final year of observing season of UID 30901}
\label{tab:slope_diff}
\begin{tabular}{@{}lcccc}
\hline
Band  & Linear phase	&     Last year	   & $\Delta$ & $\sigma$ \\
      & (mag/100 days)  &   (mag/100 days) &    (mag) & (mag)\\
\hline
\hline
z	&    0.9(0.1)	&   0.5(0.3)	 & 0.4(0.3)	& 1.3 \\
J	&    0.72(0.04)	&   0.1(0.2)	 & 0.6(0.2)	& 2.7 \\

\hline
\end{tabular}
\end{table}

During the linear decline phase, the $z$-band shows a decline rate of 0.89
$\pm$ 0.11 mag/100 days after which it drops to 0.53 $\pm$ 0.26 mag/100 days
in the final year, as determined from two points seen in the late light curve.
The difference in the decline of the $J$-band is even higher.  While in the
linear phase the $J$-band declines at 0.72 $\pm$ 0.04 mag/100 days, in the
final year it drops to 0.13 $\pm$ 0.22 mag/100 days.  These changes correspond
to a decrease in the decline rates of 0.36 $\pm$ 0.28 mag/100 days and 0.59
$\pm$ 0.22 mag/100 days, at a 1.3 and 2.7 sigma level in the $z$ and $J$
bands, respectively.  In Section \ref{PISN} we show that the late decline
rates are not consistent with the $^{56}$Ni radioactive decay. We summarise
our results in Table \ref{tab:late_decline}.

\begin{table}
\centering
\caption{Individual slope measured for individual $gizyJ$ bands}
\begin{tabular}{lccc}
\hline
Band & $\alpha$ & $n_{obs}$\\
\hline
\hline
g &  -2.8  & 5 \\
i &  -2.7  & 7 \\
z &  -3.2  & 8 \\
Y &  -3.3  & 22\\
J &  -3.0  & 17\\
\hline
\label{tab:tail_plaw}
\end{tabular}
\end{table}

A different approach to assess the potential flattening in the
light curves is the difference between the photometric 
observations in the final year and the extrapolation of the 
linear model fitted to the second observing season. These
differences yield consistently negative residuals summarised 
in Table \ref{tab:mag_diff}
which means that the SN remains consistently brighter in all bands
and at all epochs in the final year when compared with the brightness
expected from the linear extrapolation. It is important to notice
that: 1) the linear fits are based on several observations, 
therefore the parameters and their errors are considered robust,
and 2) the majority of the core collapse and thermonuclear SNe,
powered by radioactive decay are well modelled by a linear decline 
rate from about a hundred days to several hundred days after the 
explosion. 

The fact that all bands ($gizy$ and $J$) at all late epochs exhibit 
brighter magnitudes than the ones predicted by the linear 
extrapolation implies that this is not an artefact produced by 
random errors in the photometry. Additionally, the $z$ and $J$ bands 
which have more than one epoch in the last observing season show 
a decrease in their residuals (increase in absolute value) with 
time. The residual in $z$ decreases from -0.6 $\pm$ 0.4 mag to -0.8 
$\pm$ 0.5 mag over 57 days in the observed frame. Similarly, the 
$J$ band residuals decrease from -0.01 $\pm$ 0.18 mag to -0.52 
$\pm$ 0.23 mag over nearly 84 days, suggesting a pronounced 
flattening and deviation with time from the linear decline model. 
The uncertainties here correspond to the sum in quadrature of the 
photometric uncertainties and the linear model uncertainties, the 
latter being the dominant source of uncertainty. 

The apparent flattening is remarkably pronounced in $g$-band, 
where the measured residual is -2.0 $\pm$ 0.3 mag or a 6.6 sigma 
deviation from the extrapolated linear decline model. There is an 
apparent tendency for bluer bands to have a more pronounced deviation 
from the linear extrapolation compared to redder filters at similar 
epochs. The notable deviation from the linear decline measured 
in $g$-band via these residuals could be, in part, explained by the 
large decline rate observed over the second observing season, 
which predicts a much faster dimming of the SN in this band 
compared to others.

Other SLSN light curves have also been observed to flatten at late epochs. 
SN 2015bn \citep{nicholl15bn} and SN 2016inl \citep{Blanchard21} exhibit a 
flattening similar to UID 30901 and a decline rate significantly slower than 
the radioactive decay of $^{56}$Ni. In order to compare the three objects we 
fit a power law to the last two sections of the individual $gizy$ and $J$ 
bands finding values that range from $L_{i} \propto t^{-2.8}$ to $L_{Y} \propto t^{-3.3}$ (see Table \ref{tab:tail_plaw}). We also measure a $L_{r} \propto t^{-4.0}$ for SN 2015bn at the same rest-frame phase assuming $z = 0.37$, while \cite{Blanchard21} report a decline of $L \propto t^{-2.8}$ in the combined $r+$F625W filters for SN 2016inl found at $z=0.31$. In summary, UID 30901 and SN 2016inl show a similar degree of flatness in their late-time evolution, while both SLSNe have shallower power-law indices than SN 2015bn.

\subsection{Blackbody Model}
\label{sec:bol_lumin}

Armed with the light curve interpolations described in Section 
\ref{sec:lc_interpolation}, we now proceed to fit a blackbody model. 

The high densities and 
temperatures that prevail in the early stages of the explosion make 
the blackbody a good fit to the data and useful to estimate basic 
physical parameters, such as the bolometric temperature
($T_{bb}$) and radius ($R_{bb}$) of the SN, regardless of the
mechanism behind the event.

\begin{figure}
\includegraphics[scale=0.18]{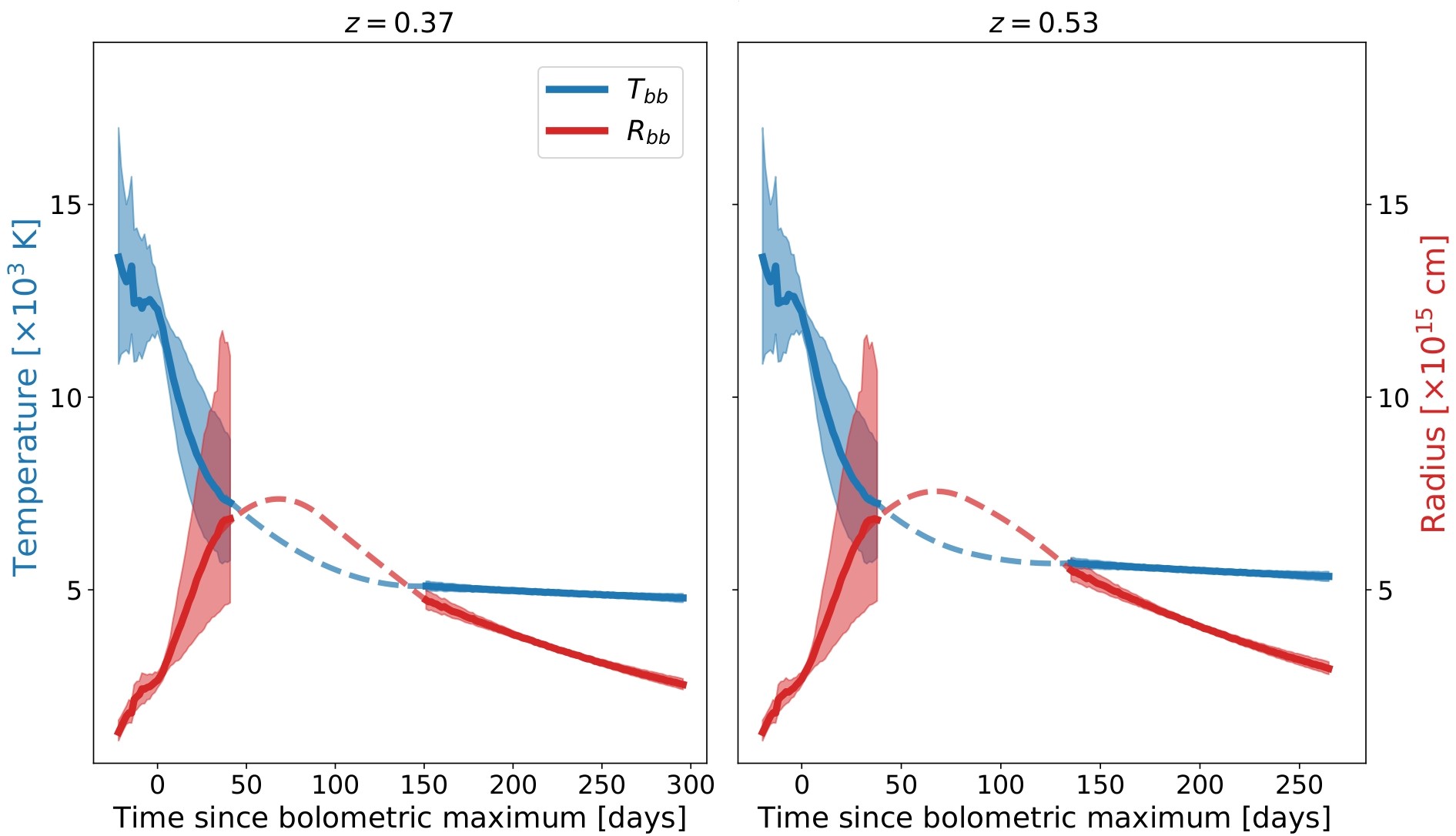}
\caption{Blackbody temperature ($T_{bb}$; in blue) and radius ($R_{bb}$; in
  red) as a function of rest-phase time since bolometric maximum. We present
  these parameters from our blackbody fits to the interpolated photometry for
  the redshifts equal to $0.37$ and $0.53$. The dashed lines represent an
  interpolation of $T_{bb}$ and $R_{bb}$.}
\label{fig:bb_params}
\end{figure}

\begin{figure}
\includegraphics[scale=0.16]{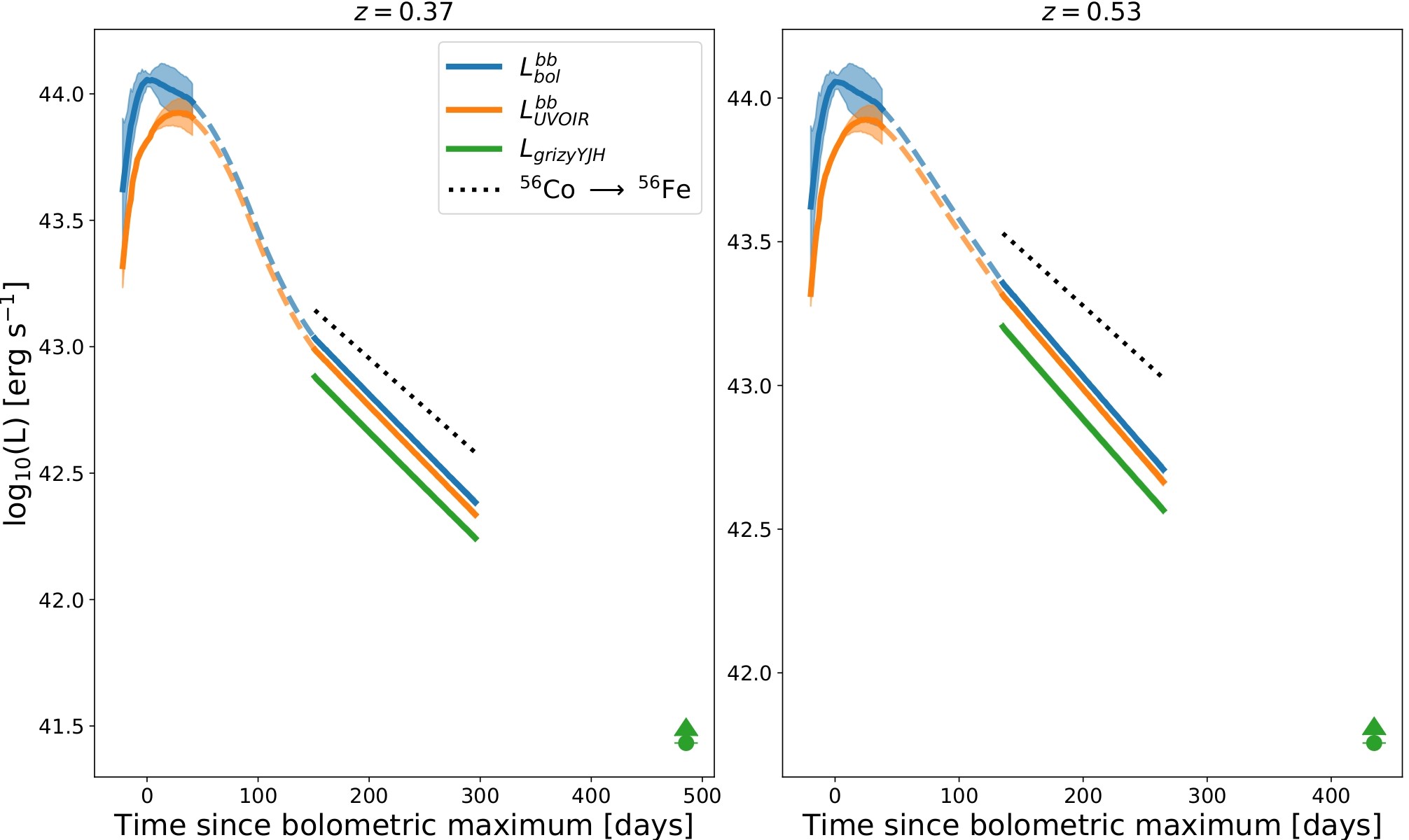}
\caption{Bolometric luminosity ($L_{\mathrm{bol}}$; in blue), UVOIR
  pseudo-bolometric luminosity ($L^{bb}_{\mathrm{UVOIR}}$; in orange), and the
  pseudo-bolometric luminosity ($L_{\mathrm{grizyYJH}}$; in green) obtained
  from summing over the luminosity emitted in the observed bands, at redshifts
  equal to $0.37$ and $0.53$. The blue and orange dashed lines correspond to
  $L_{\mathrm{bol}}$ and $L^{bb}_{\mathrm{UVOIR}}$ computed using the
  interpolated values of $T_{bb}$ and $R_{bb}$. The black-dotted lines
  correspond to the $^{56}$Co decay.  }
\label{fig:bolometric_lc}
\end{figure}

In the early phase we only interpolate
observations and do not perform any extrapolations. In this 
phase, we always have observations in at least three bands to fit a 
blackbody model. Typically we have seven bands from $g$ to $K_s$
bands, providing an excellent leverage to get reliable blackbody
parameters. 

For the linear decline phase, we interpolate and extrapolate
the light curves from MJD=56970.0 to MJD=57168.0, as described
in Section \ref{sec:linear_interpol}. During this phase,
the blackbody model does not provide a precise description of the observations.
But this is expected as SLSN spectral energy distributions (SEDs) is dominated 
by absorption and emission lines deviating from a perfect blackbody. 
However, the blackbody fits provide a good approximation to the 
pseudo-photospheric temperature and serve to estimate the total bolometric 
emission. 

For the late phase, we estimate a single pseudo-bolometric data point at
MJD = 57428.0 computed from summing the flux over the $gizy$ and
$J$ bands. To do this we used observations obtained close in time
for the $giy$ bands, and linearly interpolated the $z$ and $J$ bands, since
these two bands have several individual observations over this light
curve phase. 

To perform our blackbody fits, we assumed host galaxy redshifts of $0.37$, and
$0.53$ as described in Section \ref{sec:host}. In Figure \ref{fig:bb_params}
we present $T_{bb}$ and $R_{bb}$ and in Figure \ref{fig:bolometric_lc} we
present the bolometric luminosity ($L_{\mathrm{bol}}$), the UVOIR
pseudo-bolometric luminosity ($L^{bb}_{\mathrm{UVOIR}}$) computed by
integrating the blackbody emission from $3000$ to $20000$ \AA~ in the
rest-frame, and the pseudo-bolometric luminosity ($L_{\mathrm{grizyYJH}}$)
obtained from summing over the luminosity emitted in the observed bands, at
these redshifts. 

\subsubsection{UV absorption}
\label{sec:UVabsorption}

\begin{figure*}
    \centering
    \includegraphics[scale=0.5,trim=0 60 0 100]{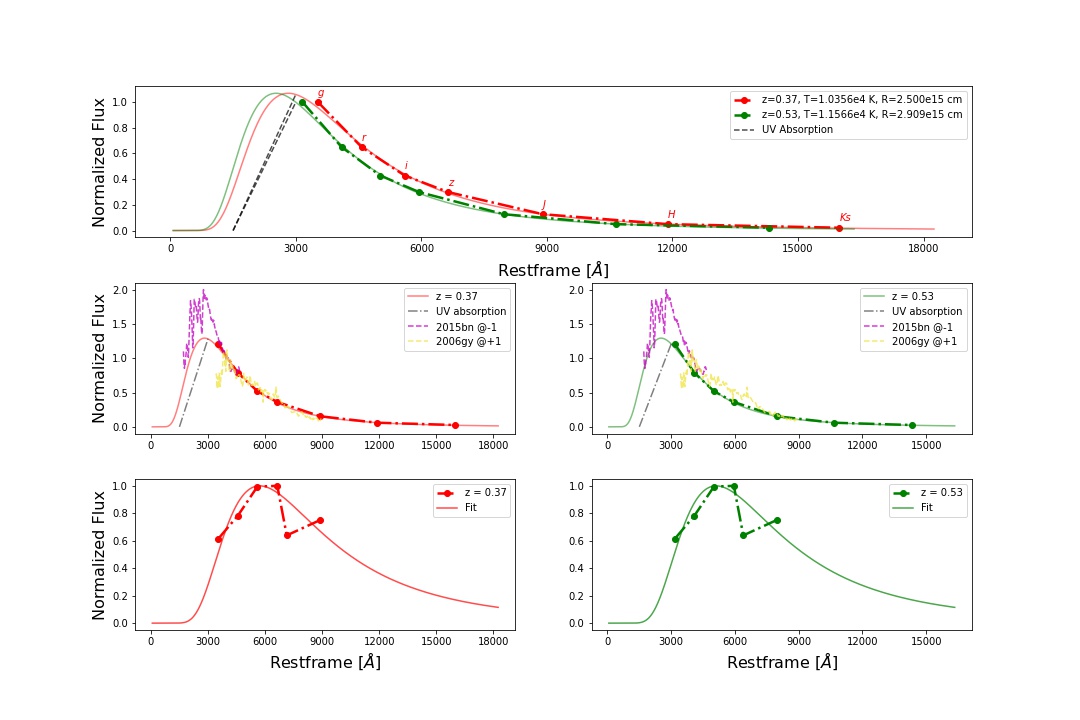}
        \caption{Comparison of the observed $grizJHKs$ SED one day after
          maximum light with blackbody models for our two tentative
          redshifts. Radii and temperature correspond to the best fit
          parameters. The black dotted line represents the linear absorption
          between 1500-3000 \AA\ expected for a SLSN. The middle panels
          compare our observations and modelling with two SLSNe, SN 2015bn and
          SN 2006gy. The bottom panels present an SED at MJD = 57035.25
          corresponding to the linear decay phase. It is clear that the
          blackbody assumption is not completely valid at these epochs.}
    \label{fig:BBatPeak}
\end{figure*}

One of the main spectral features of SLSNe is the significant absorption in
the UV part of the SED at the pre-peak phase \citep{quimby11}.  Recently
\cite{Yan18_farUV} present four UV spectra from different SLSNe showing that
the emission at $\lambda < 2800\ $ \AA\ is considerably lower than the
blackbody model. This absorption should leave an imprint in our blackbody
models. Based on this, we examine the SED of UID 30901 at MJD$\sim$56764.0 in
Figure \ref{fig:BBatPeak}, only one day after maximum light when the blackbody
assumption is highly consistent with the observations. With detections in seven bands, from $g$ to
$K_{s}$, the photometry is similar to a low resolution spectrum.

As already mentioned in Section \ref{sec:host}, we note that for $z = 1.61$ the $g$, $r$
and $i$ bands map the rest frame UV, and the best fit parameters for the
blackbody temperature and radius are $\sim 20,000$ K and $\sim 3 \times
10^{15}$cm, respectively. While the radius is within typical values for a
SLSN, the temperature exceeds the expected theoretical values by a factor of
$\sim 2$. In fact, the peak of the blackbody is found at $\sim 1500$ \AA\ and
therefore $\sim$ 73\% of the radiated energy is emitted below $3000$ \AA, in
contrast with what is expected if significant UV absorption is present.

\begin{table*}
\centering
\begin{center}
\caption{Bolometric information for UID 30901.}
\label{tab:bolometric_summary}
\begin{tabular}{@{}lccccc}
\hline
$z$      & MJD peak bolometric & Peak bolometric luminosity                      & Bolometric linear decline rate       & Total bolometric energy             & Pseudo-bolometric luminosity  \\
         & (days)              & (erg s$^{-1}$)                                  & (mag/100 days)                      & emitted (erg)                       & at MJD=57428.0 (erg s$^{-1}$) \\
\hline
$0.37$ & $56763.0$ & $5.4(3)\times 10^{43}$& $ 1.12(01) $ & $5.51 \times 10^{50}$ & $2.7(2) \times 10^{41}$ \\
$0.53$ & $56763.0$ & $1.1(1)\times 10^{44}$& $ 1.25(01) $ & $1.04 \times 10^{51}$ & $5.7(4) \times 10^{41}$\\

\hline
\end{tabular}
\begin{tablenotes}
\item Numbers in parenthesis correspond to 1-$\sigma$ statistical uncertainties.
\end{tablenotes}
\end{center}
\end{table*}

\subsection{Parametric Engine Models}\label{sec:3.5}

In this section we implement two models that may be able to account
for the high constrained energy of UID 30901: a spinning magnetar and the radioactive 
decay of $^{56}$Ni. 

\subsubsection{Power injection from the spin-down of a magnetar}

First developed by \citet{maeda07}, \citet{woosley10} and \citet{kasen10},
a rapidly rotating magnetar has been applied to Hydrogen-poor SLSNe 
providing a good fit to their light curves. Essentially,
this model consists in the collapse of a massive star creating a rapidly
spinning neutron star (P $\sim$ few milliseconds) with a strong
magnetic field ($\sim$ 10$^{14}$G). The magnetic dipole will
decay in days or weeks emitting enough high-energy radiation
to heat the ejecta and power its observed luminosity. 
The model has been modified to account for the diversity
of SLSN light curves \citep{inserra13, wang15, nicholl17}.

The principal parameters of the magnetar model are: 1) the period of the 
magnetar and its magnetic field, which control the energy input of the 
SLSN; 2) the opacity and the high-energy opacity, which control the
internal diffusion time and the high-energy "leakage" parameter
\citep{wang15}; 3) the ejecta mass and the photospheric velocity,
which control the kinetic energy and the radius of the 
photosphere; and 4) the final temperature and time of explosion. 
The evolution of the radius and temperature are controlled 
by simple rules (See Appendix \ref{sec:appendixmagnetar}) 
and the SED is assumed to be a blackbody. The model 
does not account for the host galaxy reddening or the 
\textit{modified} part of the blackbody. We implemented the magnetar 
model as described in \citet{nicholl17} and fitted it directly to the observed 
magnitudes. We fit our model with the \emph{emcee} sampler 
\citep{foreman13}. Further details of our model can be seen in the Appendix     
\ref{sec:appendixmagnetar}.

\begin{figure*}
\includegraphics[width=\textwidth]{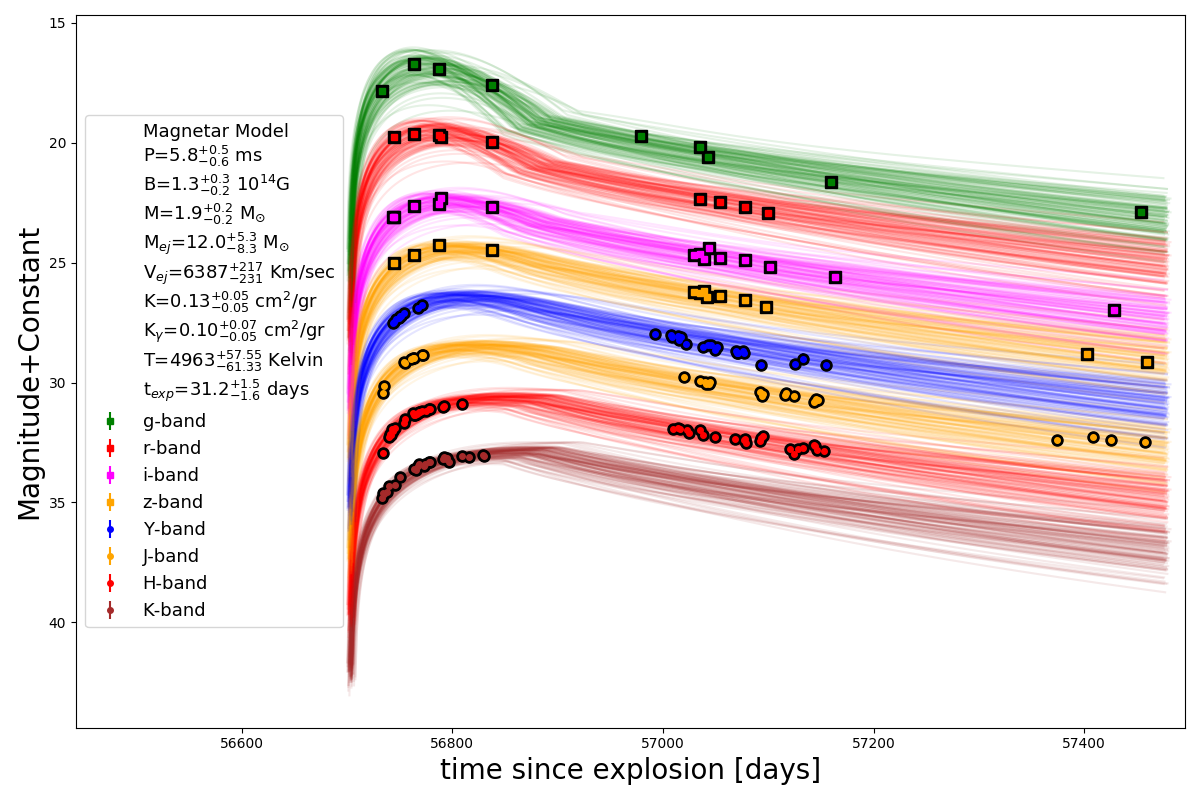}
\caption{Magnetar model fit for $z = 0.37$.}
\label{fig:mosfit}
\end{figure*}

Our model fit for $z = 0.37$ predicts a period spin
$P = 5.8^{+0.5}_{-0.6}$ms, a magnetic field 
$B = 1.3^{+0.3}_{-0.2}\times 10^{14}$G
and an $M_{\mathrm{ej}} = 12.0^{+5.3}_{-8.3}$ M$_{\odot}$.
We also find the neutron star mass $M_{\mathrm{ns}} = 1.9 \pm 0.2$ M$_{\odot}$
and an ejecta velocity of $V_{\mathrm{ej}} = 6387^{+217}_{-231} $km s$^{-1}$.
For simplicity we consider a constant 
expansion of the ejecta and this velocity is consistent
with the plateau of velocity at the tail of the light 
curves derived from the blackbody model.
We also fit the opacity $\kappa$ and the opacity to
high-energy photons $\kappa_{\gamma}$ parameters, that control the 
injection of energy from the magnetar, 
resulting in  $\kappa = 0.13 \pm 0.05 $ cm$^{2}$ gr$^{-1}$ and 
$\kappa_{\gamma} = 0.10^{+0.07}_{-0.05} $ cm$^{2}$ gr$^{-1}$. 
The model predicts an explosion time $t_0 = 31.2^{+1.5}_{-1.6}$ 
days before the first detection, which is in disagreement
with the upper limits found from non-detections. The outcome of our model 
for $z = 0.37$ is shown in Figure \ref{fig:mosfit}.

\begin{figure}
\centering
\includegraphics[width=8.5cm]{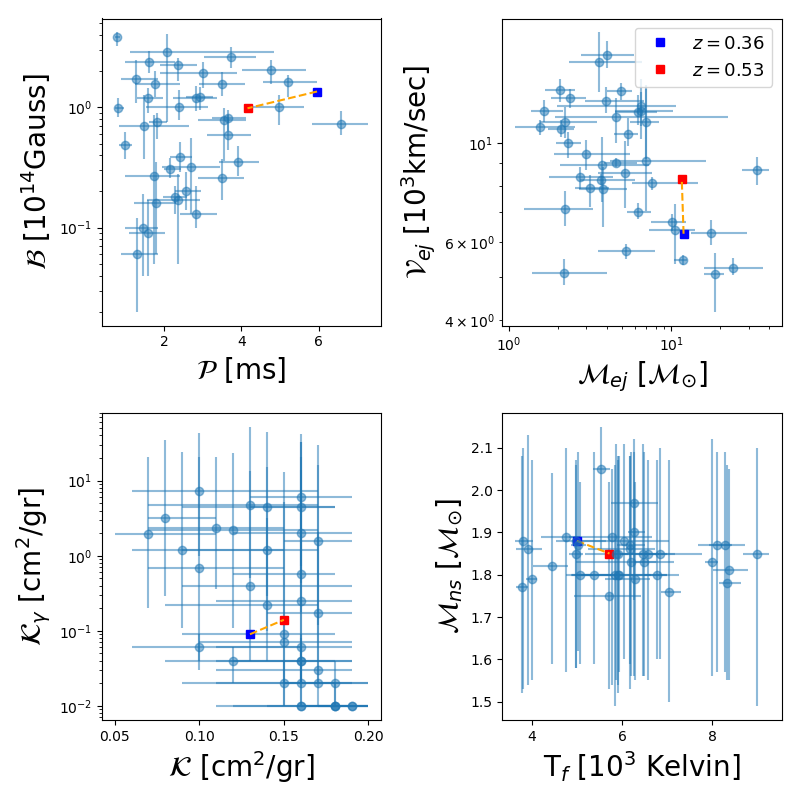}
\caption{Comparison against the parameters measured by \citet{nicholl17}. 
The blue and red squares represent the UID 30901 at $z=0.37$ and $z = 
0.53$. Our object present typical SLSNe values for both redshift. The 
period in the case of $z = 0.37$ adopt an extreme value but still 
consistent with others SLSNe}
\label{fig:nicholl17}
\end{figure}

A similar set of parameters are found for $z = 0.53$. 
We notice that a higher $z$ implies hotter and 
faster ejecta, a faster spinning magnetar and a 
weaker magnetic field. Both set of parameters are 
consistent with the typical values calculated by \citet{nicholl17} 
for a sample of 38 SLSNe-I (See Figure \ref{fig:nicholl17}).
A full set of parameters for both redshifts is presented
in Table \ref{Table: Mosfit}. 

\subsubsection{Radioactive Decay of $^{56}$Ni}
\label{PISN}

Another theoretical mechanism to reproduce the extreme 
luminosities of SLSNe is through the radioactive decay of several
solar masses of $^{56}$Ni. The main parameters of this model are 
$M_{\mathrm{ej}}$ and $M_{\mathrm{Ni}}$, the masses of the ejecta and 
$^{56}$Ni, respectively. Additional parameters include the ejecta velocity 
($V_{\mathrm{ej}}$), final temperature ($T_{f}$) and the opacity ($\kappa$). 

Our model predicts an ejecta mass of $M_{\mathrm{ej}}=5.8^{+3.3}_{-3.5}$ 
and a $^{56}$Ni mass of $M_{\mathrm{Ni}} = 4.3\pm0.1$M$_{\odot}$, which is 
beyond the nickel mass to total mass ratio found by \citet{Umeda08}.We estimate 
a final temperature $T_{f} = 4911_{-62}^{+65}$ K which is fairly consistent with 
theoretical estimations to the photospheric temperature at this phase 
\citep{dessart12}. The resulting opacity is $\kappa = 0.12^{+0.05}_{-0.07}$ 
cm$^2$ gr$^{-1}$, which is among the expected values for a SLSN powered by this 
model. This model predicts an explosion time $t_0 = 24^{+1.4}_{-1.5}$ days 
before the first detection, inconsistent with our non-detection point. 

For $z = 0.53$ our best fit predicts a $M_{\mathrm{Ni}}$ higher than
the $M_{\mathrm{ej}}$, which is nonphysical. The model also implies a 
faster and hotter ejecta and an explosion time of $\sim 21.6$
days before explosion. The full set of parameters for both redshifts 
can be found in Table \ref{Table: Mosfit}.

\bgroup
    \def\arraystretch{1.5}%
    \begin{table}
        \begin{center}
        \begin{tabular}{|l|r|r|r|r|r|r|}
        \hline
        Model&  \multicolumn{2}{|c|}{Magnetar}     &      \multicolumn{2}{|c|}{$^{56}$Ni}   \\ 
        \hline
        $z$&0.37&0.53&0.37&0.53\\
        \hline
        $\mathcal{M}_{^{56}Ni}$ [$M_{\odot}$] & --                     & --                     & 4.3$_{-0.1}^{+0.1}$    & 7.9$_{-0.4}^{+0.3}$   \\ 
        
        $\mathcal{P}$ [$ms$]                  & 5.8$_{-0.6}^{+0.5}$    & 4.2$_{-0.4}^{+0.4}$    & --                     & --                    \\ 
        
        $\mathcal{B}$    [10$^{14}$ Gauss]    & 1.3$_{-0.2}^{+0.3}$    & 1.0$_{-0.2}^{+0.2}$    & --                     & --                    \\ 
        
        $\mathcal{M}_{ns}$ [$M_{\odot}$]      & 1.9$_{-0.2}^{+0.2}$    & 1.9$_{-0.2}^{+0.2}$    & --                     & --                    \\ 
        
        $\mathcal{M}_{ej}$ [$M_{\odot}$]      & 12.0$_{-8.3}^{+5.3}$   & 11.6$_{-4.7}^{+4.5}$   & 5.8$_{-3.5}^{+3.3}$    & 6.3$_{-4.4}^{+3.2}$   \\ 
        
        $\mathcal{V}_{ej}$ [$km/s$]           & 6387$_{-231}^{+217}$   & 8303$_{-293}^{+226}$   & 7130$_{-240}^{+262}$   & 10310$_{-392}^{+561}$ \\ 
        
        $\mathcal{K}$ [$cm^2/gr$]             & 0.13$_{-0.05}^{+0.05}$ & 0.15$_{-0.03}^{+0.04}$ & 0.12$_{-0.07}^{+0.05}$ & 0.10$_{-0.06}^{+0.04}$\\ 
        
        $\mathcal{K}_{\gamma}$ [$cm^2/gr$]    & 0.10$_{-0.05}^{+0.07}$ & 0.14$_{-0.05}^{+0.06}$ & -- & -- \\ 
        
        $T$ [$Kelvin$]                        & 4963$_{-61}^{+57}$     & 5719$_{-51}^{+72}$     & 4911$_{-62}^{+65}$     & 5549$_{-91}^{+80}$    \\ 
        
        $t_{exp}$ [$days$]      & 31.2$_{-1.6}^{+1.5}$   & 31.4$_{-1.6}^{+1.6}$   & 24.0$_{-1.5}^{+1.4}$   & 21.6$_{-1.7}^{+1.5}$  \\

        \hline
        \end{tabular}
        \caption{Magnetar and $^{56}$Ni Model Fit Results. The $t_{exp}$ is reported in observer frame}
        \label{Table: Mosfit}
        \end{center}
    \end{table}
\egroup

\section{Discussion}\label{sec:discussion}

We divide this section in two parts. The first one
presents a detailed analysis of the outcome parameters of our 
model fits for $z = 0.37$, and how it changes when we assume 
$z = 0.53$. Then we discuss the explosion epoch and the possibility
that ejecta-CSM interaction as a power source for UID 30901.

In the second part of this section we analyse the light-time light curve
flattening from two perspectives, first as a power source effect
and then, as a light echo. 

\subsection{Fitting results}

\subsubsection{Power Source}
\label{sec:PowerSources}

In Section \ref{sec:3.5} we presented two models as the potential
power source for UID 30901, radioactive decay of $^{56}$Ni and the 
spin-down of a magnetar. Here we discuss the parameters resulting from 
these fits.

Explosions powered by the radioactive decay of $^{56}$Ni are expected to dim
at a rate of $\sim 1$ mag/100 days. Comparing this with the decay calculated in
Sections \ref{sec:linear_interpol} and \ref{sec:FinalYearandFlattening}, we
note a difference in factors of 1.1 for $z = 0.37$ and 1.5 for $z=0.53$
between the expected behaviour and the photometry in the final year. (See
Table \ref{tab:bolometric_summary} and Figure \ref{fig:bolometric_lc}).
Ignoring the final year, the light curve declines faster than the expected
$^{56}$Ni radioactive decay but is still consistent with this kind of
explosion (See Table \ref{tab:bolometric_summary}). In fact, this model
provides an excellent fit to first phases of the light curve.

For $z = 0.37$ our model predicts that the $M_{\mathrm{Ni}}$ is $\sim$75\%
of the $M_{\mathrm{ej}}$, which is not consistent with theoretical studies
\citep{Umeda08} or any other observation reported in the literature
\citep[e,g.][]{jerkstrand16}. For $z=0.53$ the 
$M_{\mathrm{Ni}}$ necessary to power the UID 30901 light curve is higher 
than the $M_{\mathrm{ej}}$. These very high $M_{\mathrm{Ni}}$ to 
$M_{\mathrm{ej}}$ ratios is the main reason to discard this model. 

\citet{dessart12} present a model that predicts that the photosphere 
of a PISN is essentially cold, reaching temperatures of 6000 K at 
the peak. We derived a peak blackbody temperature ($T_{bb}\sim12000$ K)
which is not compatible with those predictions.
At latter phases a similar behaviour is found. While the model 
predicts a maximum temperature of $\sim 4000$ K between 100-200 
days after the peak, we estimate a temperature of $\sim 5000$ K at 
the same epochs. Nevertheless, it is important to mention that the
blackbody modelling is not completely reliable at late epochs 
of the light curve.

The unphysically large values of $M_{\mathrm{Ni}}$, the inconsistent 
temperature predictions, and the difficulty of this model 
to reproduce the flattening strongly argue against $^{56}$Ni decay as 
the main power source for UID 30901.

The second scenario is the spin-down of a magnetar. This model 
reproduces the complete light curve evolution and the best fit 
parameters are in good agreement with the values typically
found in the literature. 

From the first detection to $\sim$ 200 days, the resulting bolometric 
luminosity and temperatures values derived from the blackbody fit are
consistent with those determined from the spin-down model. The main 
difference is the temperature at the nebular phase, where the resulting
temperatures differ by 20\%. This is expected since the blackbody 
assumption is no longer valid at these late epochs. 

Our model predicts $M_{\mathrm{ej}} = 11.9^{+4.8}_{-6.4}$ M$_{\odot}$, 
$B =1.4^{+0.3}_{-0.3} $ G and $P = 6.0^{+0.6}_{-0.5}$ ms.
These results are among the expected values for a SLSNe 
(See Figure \ref{fig:nicholl17}). 
For $z = 0.53$ the model yields in a weaker magnetic field,
a slower spin period and a similar ejecta mass. It also gives
a faster and hotter ejecta. The neutron star mass remains constant 
for both redshifts. A complete set of parameters is shown
in Table \ref{Table: Mosfit}

\subsubsection{Explosion time}

One of the parameters of our models is the explosion time t$_{0}$. We find
that both the radioactive decay and the magnetar model predict explosion times
in disagreement with the observations. The results are presented in Table
\ref{Table: Mosfit}.

The magnetar model has shown mixed results when estimating explosion times for
other systems. While for SN 2015bn \citep{nicholl2016} and SN 2016iet
\citep{Gomez2019} discrepancies between the observations and the explosion
time are found, for SN 2017cgi \citep{Fiore2021} fits a rise time consistent
with observations and also with a fast-evolving SLSNe.

\subsubsection{Circumstellar Interaction}
\label{secWACSM}

Another potential scenario to explain the brightness of a SLSNe
is the interaction between the SN ejecta with the CSM. 
The progenitor may be surrounded by CSM ejected from the progenitor itself 
before the explosion. Afterwards, the SN ejecta will collide with this material, 
and by converting the kinetic energy of the ejecta into radiative energy, 
can power the luminosity of the SN. This interaction can be responsible 
for bumps or wiggles in the light curves of some SLSNe \citep{yan15}.

Unlike the double-peaked light curve of iPTF13ehe \citep{yan15}, the UID 30901
light curve does not show any strong observational signature of the
interaction between the ejecta and CSM. 


Spectral features are the main evidence for interaction with the CSM, as seen
in SN\,2006gy \citep{jerkstrand20}. Because of the lack of spectra at the
photospheric phase, we cannot test for the presence of any H$\alpha$ broad
line to confirm the ejecta-CSM interaction. However, a strong interaction
should leave an imprint in the SED, visible as a deviation from the blackbody
model. A close inspection of the SED at individual epochs of the early and
post peak phases (see Figure \ref{fig:BBatPeak}) shows that the photometry is
in good agreement with a blackbody model, and therefore the presence of a
strong H$\alpha$ broad line is unlikely. In Section \ref{sec:3.3.2} we found that the
late-time SED of UID 30901 looks very similar to that of SN\,2018bsz, which
hints at the presence of CSM interaction at these phases. However, this late
event clearly cannot be responsible for the initial power injection for UID
30901 but instead it would correspond to a possible late evolution.

In summary, we cannot rule out some ejecta-CSM interaction, but under 
the criteria outlined above, there is no evidence that this 
mechanism was the main power source for UID 30901. 

\subsection{Flattening}
One of the features of the UID 30901 light curves is the flattening at late
epochs  (see Section \ref{sec:FinalYearandFlattening}), which is 
especially clear in the $g$-band at +500 days after the explosion 
(see Figure \ref{fig:opnir_lc}). In this section we approach this 
phenomena from two different perspectives. The first one is associated 
to the power source. While the magnetar and CSM models could explain 
the flattening, the constant decline rate of radioactive decay of 
$^{56}$Ni cannot explain this behaviour. 

Another option to explain the flattening is a light echo. To 
study that case, we compare the color at the peak with the color 
at the nebular phase. 

\subsubsection{Flattening as a power source effect}


As is presented in Section \ref{sec:PowerSources}, the magnetar model 
can reproduce the complete light curve evolution for both possible redshifts
of UID 30901, and the outcome parameters are in good agreement with the 
literature and with those obtained from the blackbody modelling.

In the case of CSM interaction, it is more difficult to conclude.
Assuming an ejecta expanding with a velocity between $4000 - 10000$ km s$^{-1}$, 
a shell of matter located at $r \sim 1-4 \times 10^{16}$ cm could flatten 
the slope of the light curve leaving also an imprint in the spectra. 
Without spectroscopic data, we cannot discard
the existence of such interaction, but as we discuss in Section 
\ref{secWACSM} we conclude that it is unlikely for this model to be
the main power source and therefore is also unlikely to be the responsible of 
the change in light curve decay rate. 


\subsubsection{Light Echo}

Light echoes occur when light emitted at early phases of the SN is
reflected by dust and observed at later times. While electron scattering 
is achromatic, dust scattering is efficient in reflecting blue light so the color
at the echo phase should be as blue as the peak, as we are observing the 
same light, but reflected some time later in the line of sight.
Light echoes have been observed in SNe II \citep{crotss88, suntzeff88} and
also, late observations of SN 2006gy have been interpreted as a light echo \citep{Miller2010}. 

To study this possibility, we compare the color at the peak with the color at 
the nebular phase in every band where data are available ($gizYJ$), and where 
the last detections of $z$ and $J$ are the linear interpolation of the late 
epochs described in Section \ref{sec:FinalYearandFlattening}. 

We measured a color $(g-i)_p = 0.08 \pm 0.003 $ at the peak, 
which is not consistent with the red color observed in the last 
datapoint of $(g-i)_l = 1.93 \pm 0.013$. A similar case 
occurs with $(g-z)$, $(g-y)$ and $(g-J)$. A smaller difference, but 
still inconsistent when comparing the early and late photometry is 
found in the $(i-J)$, $(z-J)$ and $(y-J)$ colors.

A different scenario occurs in $izy$ bands. For $(i-z)$ we found similar 
values between the peak and the nebular phase and found a 
bluer color at the nebular phase for $(i-y)$ and $(z-y)$.
We associate this behaviour to the presence of emission lines typical
of a nebular spectra at the $izy$ bands. Without spectra at these 
epochs we cannot identify which lines might be present, but based 
on the possible values of $z$ for UID 30901, we can discard a strong 
late-H$\alpha$ emission. 

In summary, the discrepancy between the maximum light and late 
phase colours in many filter-pairs is sufficient to discard a light echo 
as the responsible mechanism to explain of the flattening of the light curves.

\begin{table}
\begin{center}
\caption{Color and color difference between peak and tail of the UID 30901 in 
different photometric filters.}
\label{tab:color}
\begin{tabular}{cccc}
\hline

Bands & Color at peak & Color at tail & $\Delta _{color}$\\
      & MJD $\sim$ 56764   & MJD $\sim$ 57450   &    \\
      & (mag) & (mag) & (mag) \\
\hline
\hline

$(g-i)$ & 0.08  (0.003)  & 1.93  (0.01) & -1.85  (0.02) \\
$(g-z)$ & 0.06  (0.02)   & 1.77  (0.06) & -1.71  (0.09) \\
$(g-y)$ & -0.16 (0.02)   & 1.45  (0.04) & -1.61  (0.01) \\
$(g-j)$ & -0.23 (0.02)   & 2.43  (0.03) & -2.66  (0.05) \\
$(i-z)$ & -0.02 (0.03)   & -0.0  (0.1)  & -0.0   (0.1) \\
$(i-y)$ & -0.25 (0.03)   & -0.48 (0.02) & 0.231  (0.002) \\
$(i-j)$ & -0.31 (0.02)   & 0.57  (0.03) & -0.88  (0.05) \\
$(z-y)$ & -0.23 (0.001)  & -0.5  (0.1)  & 0.2    (0.1) \\
$(z-j)$ & -0.29 (0.002)  & 0.52  (0.06) & -0.81  (0.06) \\
$(y-j)$ & -0.07 (0.003)  & 0.98  (0.07) & -1.05  (0.07) \\

\hline
\end{tabular}
\end{center}
\end{table}

\section{Summary} \label{sec:summary}

We have presented multi-wavelength photometry for UID 30901,
a new SLSN discovered in the COSMOS field by the NIR UltraVISTA SN 
Survey. Even though the UltraVISTA project was not aimed for the search of 
transients, its depth and high cadence plus the high quality of the achieved 
photometry make this survey optimal for a transient search. 

We complement the NIR data with photometry from DECam ($griz$)
and SUBARU-HSC ($grizy$). This allows us to have a wide wavelength
coverage that makes the UID 30901 one of the best observed SLSN to date.
This wide coverage let us compare the photometry with spectra available in the 
literature. The data show that UID 30901 belongs to the "sub-luminous" 
family of SLSNe, reaching a peak apparent magnitude of -20 or more in all 
observed bands, making it hard to be observed by other surveys such as PTF.

Analysis of the light curves show that the object peaks
later at longer effective wavelengths, similar to SNe Ic and Ic-BL. 
Similarities between these two kind of objects has been seen before,
finding that even though they are different classes, there is an overlap in
their peak magnitudes among the brighter Ic and the low luminosity SLSNe.

To determine the explosion time we applied two simple fits to the rise 
phase of the light curves, a linear polynomial and a power law. The last 
one provides an explosion time constraint which differs by $\sim 10$ days
with the linear fit but which is in better agreement with a non-detection
in the $H$-band five days before the fist detection of the SN. We note that 
the NIR light curve shape at rise phase can be reproduced by a power law 
with $\alpha \sim 0.7$, an odd value compared with the typical $\alpha$ 
found for other SNe. However, there are no systematic studies that explore 
SLSNe in the NIR in order to establish a correlation between the 
rise phase and power law fits. 

We detect the host of UID 30901 in a deep HST F814W image. The host is identified
as a faint, dwarf galaxy most likely characterised by a small stellar mass
and low SFR, placing it at the extreme of the properties of SLSN hosts, as 
determined by \cite{perley16}. Since we have no spectra for neither the host 
nor the SN, we adopt the photometric-$z$ values provided by the COSMOS2015
catalog for possible nearby companions, identified as galaxies A and B, with 
values $z$ = 0.37 and $z$ = 0.53, respectively.

Despite the lack of spectra, we were able to reconstruct the SEDs for 
individual epochs at different phases of the light curve evolution.
The early phases yielded the most reliable SEDs. Following the assumption that
SLSNe radiate as a blackbody we estimated the radius and temperature
evolution and a bolometric and pseudo-bolometric light curves.
Since it is known that below $\sim 3000$ \AA\, the SED of a 
SLSN suffers from significant absorption and our blackbody modelling 
for $z > 1$ implies that most of the emission would have occurred 
in the near and far UV, we were able to discard a high $z$ value for
UID 30901.

We explored the most common power sources for SLSNe and applied these
physical models to our source. The high $^{56}$Ni masses necessary to power
this SLSNe and, to a lesser extent, the estimated high temperatures at the 
peak lead us to discard the radioactive decay as the main power source
for the UID 30901. 

We consider that there is no strong physical reasons to model 
this object as an ejecta-CSM interacting SLSN. The smoothness
of the light curve, absence of early bumps and the lack of signs
of interaction of the individual SEDs makes it very unlikely 
for this option to be the main power source. 

In the case of the spin-down magnetar model, we found that
it provides an excellent fit and typical values for all parameters.
Our model predicts a magnetic field $B \approx 1.3 \times 10^{14}$ G,
a period spin $P \approx 5.8$ ms and ejecta mass of 
$ M_{\mathrm{ej}} \approx 12.0$ M$_{\sun}$ for $z=0.37$. For $z = 0.53$ we 
found $B \approx 1.0 \times 10^{14}$ G, $P \approx 4.2$ ms, M$_{ej}
\approx 11.6$ M$_{\odot}$. In both cases these physical parameters 
place our object as a typical SLSNe. The main drawback of the 
magnetar model is that it overestimates the explosion time $t_0$,
in contradiction with observed constraints.

\section{Acknowledgements} \label{sec:aknowledgements}

Based on data products from observations made with ESO Telescopes at 
the La Silla Paranal Observatory under ESO programme ID 179.A-2005 and 
on data products produced by CALET and the Cambridge Astronomy Survey 
Unit on behalf of the UltraVISTA consortium.  

This research uses services or data provided by the Astro Data Archive 
at NSF's National Optical-Infrared Astronomy Research Laboratory. NSF's
OIR Lab is operated by the Association of Universities for Research in 
Astronomy (AURA), Inc. under a cooperative agreement with the National 
Science Foundation.

\section*{Data Availability}

The photometric data of UID 30901 is included in the Appendix
\ref{sec:appendix_phot} of this article. It also can be found
in the online supplementary material and is available on the 
CDS VizieR facility.



\bibliographystyle{mnras}
\bibliography{UID30901} 

\appendix
\label{sec:appendix}

\section{Magnetar Fit}
\label{sec:appendixmagnetar}

Our simplified Magnetar model is based on \citet{nicholl17}'s model.
Where the following quantities are defined as:

\begin{equation}
    E_{mag} = \frac{1}{2}I\omega^2 = 2.6\times 10^{52}\left(\frac{M_{NS}}{1.4 
    M_{\odot}}\right)^{\frac{3}{2}} \left(\frac{P}{1 ms}\right)^{-2} erg
\end{equation}

is the rotational energy available from the neutron star.

\begin{equation}
    t_{mag}\approx\frac{P}{2\dot{P}}=1.3\times 10^5 \left(\frac{M_{NS}}{1.4 
    M_{\odot}}\right)^{\frac{3}{2}}\left(\frac{P}{1 
    ms}\right)^2\left(\frac{B_{\perp}}{10^{14}G}\right)^{-2}s
\end{equation}

is the spin-down timescale. Both expression give the dipole
radiation power:

\begin{equation}
    F(t) = \frac{E_{mag}}{t_{mag}}\frac{1}{(1-t/t_{mag})^2}
\end{equation}

This power is input into the ejecta where most of it is thermalized
and re-emitted by the SN as derived by \citet{arnett82}:

\begin{equation}
    L(t)=2e^{-(t/t_{diff})^2} (1- e^{-A/t^2})\int_0^t 
    F(t')\left(\frac{t'}{t_{diff}}\right)e^{(t'/t_{diff})^2} \frac{dt'}{t_{diff}} 
\end{equation}

where the diffusion time $t_{diff}$ is:

\begin{equation}
    t_{diff}=\left(\frac{2 \kappa M_{ej}}{\beta c V_{ej}}\right)^\frac{1}{2}
\end{equation}

and the ``leakage'' parameter introduced by \citet{wang15} 
that allows some of the high energy escape through the ejecta is:

\begin{equation}
    A =\frac{3 \kappa_{\gamma} M_{ej}}{4 \pi V_{ej}^2}
\end{equation}
A set of simple and empirically motivated rules control Radius and Temperature of the 
blackbody through the time:

\begin{align}
\Delta = \left(\frac{L(t)}{4\pi\sigma v_{phot}^2 t^2}\right)^{\frac{1}{4}}\\
T_{phot}(t) =
    \begin{cases}
    \Delta & \text{if } \Delta > T_f \\
    T_f    & \text{if } \Delta \leq T_f \\
    \end{cases}\\
  R_{phot}(t) =
    \begin{cases}
      v_{phot} t & \text{if } \Delta > T_f \\
      \left(\frac{L(t)}{4\pi\sigma T_f^4}\right)^{\frac{1}{2}}    & \text{if } \Delta 
      \leq T_f \\
    \end{cases}
\end{align}

where the $T_f$ is the final temperature reached by the blackbody, typically between 
4000 and 7000 K \citep{nicholl17}.
From the blackbody Radius and Temperature the photometry is estimated for the different 
bandpasses (considering the redshift)
and compared to the observed photometry. We simultaneously fitted all the model's 
parameters maximizing the likelihood $\mathcal{L}$:

\begin{equation}
ln (\mathcal{L}) = - \frac{1}{2} \sum{\left[\frac{O_i-M_i}{\sigma_i^2 +\sigma^2} + 
ln(\sigma_i^2+\sigma^2)\right]}
\end{equation}

where $O_i, \sigma_i$ are the photometry values and errors, $M_i$ is the corresponding 
model value,
and $\sigma$ is an additional error value to control underestimation of the photometry 
errorbars.
To compute the best fit parameter values we use the \emph{emcee} package 
\citep{foreman13}.

\section{Photometry}
\label{sec:appendix_phot}


\begin{table*}
\caption{UltraVISTA NIR Photometry of UID 30901.}
\label{tab:nir}
\begin{tabular}{@{}cccccc}
\hline
Date UT & MJD & $Y$ & $J$ & $H$ & $K_{s}$ \\
\hline
2014-03-17 & $56733.13$ & --                & --                & --                & $22.815$($0.143$) \\
2014-03-18 & $56734.12$ & --                & $22.472$($0.061$) & $22.958$($0.143$) & $22.614$($0.099$) \\
2014-03-19 & $56735.11$ & --                & $22.158$($0.051$) & --                & $22.652$($0.159$) \\
2014-03-22 & $56738.13$ & --                & --                & --                & $22.554$($0.132$) \\
2014-03-24 & $56740.08$ & --                & --                & $22.296$($0.084$) & $22.313$($0.124$) \\
2014-03-25 & $56741.10$ & --                & --                & $22.152$($0.075$) & --                \\
2014-03-26 & $56742.04$ & --                & --                & $22.219$($0.090$) & --                \\
2014-03-27 & $56743.06$ & --                & --                & $21.960$($0.072$) & --                \\
2014-03-28 & $56744.07$ & $21.555$($0.043$) & --                & $22.038$($0.076$) & --                \\
2014-03-29 & $56745.13$ & $21.482$($0.043$) & --                & --                & --                \\
2014-03-30 & $56746.08$ & $21.385$($0.045$) & --                & $21.896$($0.103$) & $22.291$($0.113$) \\
2014-04-02 & $56749.16$ & $21.316$($0.039$) & --                & --                & --                \\
2014-04-03 & $56750.04$ & $21.250$($0.036$) & --                & --                & $21.962$($0.083$) \\
2014-04-04 & $56751.08$ & $21.232$($0.039$) & --                & --                & --                \\
2014-04-07 & $56754.08$ & $21.131$($0.038$) & $21.177$($0.043$) & $21.716$($0.068$) & --                \\
2014-04-08 & $56755.08$ & --                & $21.219$($0.042$) & $21.535$($0.043$) & --                \\
2014-04-14 & $56761.08$ & --                & $21.049$($0.040$) & --                & --                \\
2014-04-16 & $56763.15$ & --                & $21.004$($0.038$) & $21.298$($0.045$) & --                \\
2014-04-17 & $56764.06$ & --                & $20.973$($0.041$) & $21.360$($0.046$) & $21.596$($0.078$) \\
2014-04-18 & $56765.02$ & --                & --                & $21.360$($0.044$) & --                \\
2014-04-19 & $56766.04$ & --                & --                & $21.316$($0.044$) & $21.667$($0.082$) \\
2014-04-20 & $56767.05$ & $20.913$($0.044$) & --                & $21.286$($0.043$) & $21.497$($0.039$) \\
2014-04-21 & $56768.03$ & --                & --                & $21.255$($0.044$) & $21.424$($0.051$) \\
2014-04-22 & $56769.02$ & --                & --                & $21.222$($0.036$) & --                \\
2014-04-24 & $56771.07$ & $20.768$($0.038$) & $20.884$($0.039$) & $21.204$($0.036$) & --                \\
2014-04-25 & $56772.07$ & --                & $20.858$($0.040$) & --                & --                \\
2014-04-25 & $56772.96$ & --                & --                & --                & $21.491$($0.041$) \\
2014-04-27 & $56774.03$ & --                & --                & --                & $21.424$($0.039$) \\
2014-04-27 & $56774.98$ & --                & --                & $21.191$($0.050$) & --                \\
2014-05-01 & $56778.03$ & --                & --                & $21.096$($0.040$) & $21.341$($0.045$) \\
2014-05-02 & $56779.04$ & --                & --                & $21.126$($0.044$) & $21.331$($0.047$) \\
2014-05-13 & $56790.99$ & --                & --                & $21.011$($0.064$) & $21.216$($0.050$) \\
2014-05-15 & $56792.02$ & --                & --                & $21.005$($0.042$) & $21.123$($0.029$) \\
2014-05-16 & $56793.97$ & --                & --                & --                & $21.201$($0.047$) \\
2014-05-18 & $56795.01$ & --                & --                & --                & $21.142$($0.049$) \\
2014-05-19 & $56796.98$ & --                & --                & --                & $21.325$($0.042$) \\
2014-05-31 & $56808.99$ & --                & --                & $20.889$($0.049$) & $21.081$($0.046$) \\
2014-06-08 & $56816.00$ & --                & --                & --                & $21.093$($0.048$) \\
2014-06-20 & $56829.00$ & --                & --                & --                & $21.049$($0.056$) \\
2014-06-21 & $56829.97$ & --                & --                & --                & $21.066$($0.047$) \\
2014-12-01 & $56992.32$ & $21.988$($0.045$) & --                & --                & --                \\
2014-12-16 & $57007.27$ & $22.056$($0.046$) & --                & --                & --                \\
2014-12-17 & $57008.25$ & $22.131$($0.045$) & --                & --                & --                \\
2014-12-18 & $57009.27$ & --                & --                & $21.952$($0.048$) & --                \\
2014-12-23 & $57014.28$ & $22.097$($0.045$) & --                & $21.907$($0.045$) & --                \\
2014-12-24 & $57015.23$ & $22.260$($0.059$) & --                & --                & --                \\
2014-12-25 & $57016.26$ & --                & --                & $21.945$($0.045$) & --                \\
2014-12-26 & $57017.29$ & $22.136$($0.048$) & --                & --                & --                \\
2014-12-29 & $57020.22$ & --                & $21.798$($0.075$) & --                & --                \\
2014-12-31 & $57022.24$ & $22.402$($0.092$) & --                & --                & --                \\
2015-01-01 & $57023.20$ & --                & --                & $22.001$($0.087$) & --                \\
2015-01-03 & $57025.22$ & --                & --                & $22.112$($0.104$) & --                \\
2015-01-13 & $57035.31$ & --                & $21.969$($0.052$) & $21.979$($0.107$) & --                \\
2015-01-16 & $57038.29$ & $22.539$($0.063$) & --                & $22.214$($0.058$) & --                \\
2015-01-18 & $57040.31$ & --                & $22.005$($0.040$) & --                & --                \\
2015-01-19 & $57041.33$ & --                & $22.078$($0.057$) & --                & --                \\
2015-01-21 & $57043.29$ & $22.459$($0.058$) & $22.082$($0.114$) & --                & --                \\
2015-01-22 & $57044.33$ & --                & $21.986$($0.064$) & --                & --                \\
2015-01-23 & $57045.26$ & $22.441$($0.054$) & --                & --                & --                \\
2015-01-27 & $57049.22$ & $22.671$($0.077$) & --                & $22.301$($0.071$) & --                \\
2015-01-29 & $57051.18$ & $22.537$($0.052$) & --                & --                & --                \\
  \hline
 \end{tabular}
 \begin{tablenotes}
        \item Numbers in parenthesis correspond to 1\,$\sigma$ statistical uncertainties.
 \end{tablenotes}
\end{table*}

\begin{table*}
\contcaption{UltraVISTA NIR Photometry of UID 30901.}
\begin{tabular}{@{}cccccc}
\hline
Date UT & MJD & $Y$ & $J$ & $H$ & $K_{s}$ \\
\hline
2015-02-15 & $57068.18$ & --                & --                & $22.383$($0.069$) & --                \\
2015-02-16 & $57069.10$ & $22.721$($0.057$) & --                & --                & --                \\
2015-02-17 & $57070.18$ & $22.775$($0.054$) & --                & --                & --                \\
2015-02-18 & $57071.18$ & $22.779$($0.092$) & --                & --                & --                \\
2015-02-23 & $57076.14$ & $22.712$($0.049$) & --                & --                & --                \\
2015-02-24 & $57077.14$ & $22.782$($0.049$) & --                & $22.419$($0.111$) & --                \\
2015-02-25 & $57078.06$ & --                & --                & $22.373$($0.083$) & --                \\
2015-02-26 & $57079.09$ & --                & --                & $22.532$($0.136$) & --                \\
2015-03-11 & $57092.20$ & --                & $22.396$($0.117$) & $22.469$($0.104$) & --                \\
2015-03-12 & $57093.14$ & $23.296$($0.171$) & --                & $22.314$($0.077$) & --                \\
2015-03-13 & $57094.09$ & --                & $22.570$($0.114$) & --                & --                \\
2015-03-14 & $57095.06$ & --                & $22.486$($0.092$) & $22.246$($0.093$) & --                \\
2015-04-04 & $57116.09$ & --                & $22.541$($0.054$) & --                & --                \\
2015-04-05 & $57117.06$ & --                & $22.457$($0.075$) & --                & --                \\
2015-04-09 & $57121.14$ & --                & --                & $22.787$($0.174$) & --                \\
2015-04-12 & $57124.07$ & --                & $22.595$($0.073$) & $22.971$($0.159$) & --                \\
2015-04-13 & $57125.13$ & $23.252$($0.188$) & --                & --                & --                \\
2015-04-16 & $57128.03$ & --                & --                & $22.774$($0.137$) & --                \\
2015-04-21 & $57133.05$ & $23.025$($0.217$) & --                & $22.723$($0.125$) & --                \\
2015-05-01 & $57143.06$ & --                & $22.846$($0.103$) & $22.611$($0.165$) & --                \\
2015-05-02 & $57144.03$ & --                & $22.790$($0.077$) & $22.660$($0.139$) & --                \\
2015-05-03 & $57145.10$ & --                & $22.702$($0.082$) & --                & --                \\
2015-05-04 & $57146.01$ & --                & --                & $22.842$($0.125$) & --                \\
2015-05-05 & $57147.09$ & --                & $22.739$($0.116$) & --                & --                \\
2015-05-11 & $57153.09$ & --                & --                & $22.847$($0.170$) & --                \\
2015-05-13 & $57155.05$ & $23.283$($0.173$) & --                & --                & --                \\
2015-12$^{\mathrm{a}}$ & $57374.10$ & --    & $24.399$($0.127$) & --                & --                \\
2016-01$^{\mathrm{a}}$ & $57408.76$ & --    & $24.289$($0.092$) & --                & --                \\
2016-02$^{\mathrm{a}}$ & $57425.34$ & --    & $24.425$($0.156$) & --                & --                \\
2016-03$^{\mathrm{a}}$ & $57457.94$ & --    & $24.488$($0.138$) & --                & --                \\
  \hline
 \end{tabular}
 \begin{tablenotes}
        \item Numbers in parenthesis correspond to 1\,$\sigma$ statistical uncertainties.
        \item $^{\mathrm{a}}$The $J$-band photometry was computed from an average image made of several frames co-added obtained over the month.
 \end{tablenotes}
\end{table*}

\begin{table*}
\caption{Optical photometry of UID 30901}
\label{tab:optical}                                                                                                                                                                                                                        
\begin{tabular}{lccccccc}
  \hline
  \multicolumn{1}{l}{Date UT} &
  \multicolumn{1}{c}{MJD} &
  \multicolumn{1}{c}{$g$} &
  \multicolumn{1}{c}{$r$} &
  \multicolumn{1}{c}{$i$} &
  \multicolumn{1}{c}{$z$} &
  \multicolumn{1}{c}{$y$} &
  \multicolumn{1}{c}{Tel.} \\
  \hline
2014-03-17 & $56733.14$ & $21.911$($0.125$) & --                & --                & --                & --                & 1 \\
2014-03-25 & $56741.47$ & --                & --                & --                & --                & $21.547$($0.015$) & 2 \\
2014-03-26 & $56742.98$ & --                & --                & $21.120$($0.039$) & --                & $21.488$($0.013$) & 2 \\
2014-03-28 & $56744.35$ & --                & $20.922$($0.024$) & --                & $21.060$($0.018$) & --                & 2 \\
2014-03-28 & $56744.51$ & --                & --                & $21.141$($0.020$) & --                & --                & 2 \\
2014-04-03 & $56750.29$ & --                & --                & --                & --                & $21.015$($0.020$) & 2 \\
2014-04-17 & $56764.06$ & $20.790$($0.021$) & $20.679$($0.017$) & $20.679$($0.018$) & $20.688$($0.043$) & --                & 1 \\
2014-05-09 & $56786.97$ & $20.998$($0.130$) & $20.744$($0.185$) & $20.587$($0.153$) & $20.287$($0.187$) & --                & 1 \\
2014-05-11 & $56788.97$ & --                & $20.792$($0.150$) & $20.350$($0.075$) & --                & --                & 1 \\
2014-06-29 & $56837.96$ & $21.662$($0.098$) & $21.016$($0.087$) & $20.725$($0.121$) & $20.513$($0.122$) & --                & 1 \\
2014-11-18 & $56979.62$ & $23.777$($0.033$) & --                & --                & --                & $21.727$($0.015$) & 2 \\
2015-01-07 & $57029.27$ & --                & --                & $22.718$($0.087$) & $22.261$($0.093$) & --                & 1 \\
2015-01-13 & $57035.26$ & $24.251$($0.149$) & $23.404$($0.055$) & $22.691$($0.036$  & $22.305$($0.054$) & --                & 1 \\
2015-01-17 & $57039.05$ & --                & --                & $22.876$($0.023$) & $22.215$($0.046$) & --                & 2 \\
2015-01-19 & $57041.00$ & --                & --                & --                & $22.465$($0.031$) & $22.259$($0.013$) & 2 \\
2015-01-20 & $57042.46$ & $24.646$($0.020$) & --                & --                & --                & --                & 2 \\
2015-01-21 & $57043.66$ & --                & --                & $22.432$($0.070$) & --                & --                & 2 \\
2015-01-27 & $57049.65$ & --                & --                & --                & --                & $22.371$($0.015$) & 2 \\
2015-02-01 & $57054.29$ & --                & $23.502$($0.106$) & $22.853$($0.048$) & $22.436$($0.070$  & --                & 1 \\
2015-02-25 & $57078.12$ & --                & $23.735$($0.185$) & $22.935$($0.096$) & $22.568$($0.092$) & --                & 1 \\
2015-03-16 & $57097.44$ & --                & --                & --                & $22.857$($0.033$) & --                & 2 \\
2015-03-18 & $57099.45$ & --                & $23.967$($0.037$) & --                & --                & --                & 2 \\
2015-03-20 & $57101.39$ & --                & --                & $23.201$($0.070$) & --                & --                & 2 \\
2015-05-17 & $57159.31$ & $25.712$($0.078$) & --                & --                & --                & --                & 2 \\
2015-05-21 & $57163.30$ & --                & --                & $23.644$($0.066$) & --                & --                & 2 \\
2016-01-15 & $57402.67$ & --                & --                & --                & $24.852$($0.111$) & --                & 2 \\
2016-02-10 & $57428.00$ & --                & --                & $25.009$($0.183$) & --                & --                & 2 \\
2016-02-22 & $57440.00$ & --                & --                & --                & --                & $25.456$($0.207$) & 2 \\
2016-03-07 & $57454.33$ & $26.965$($0.170$) & --                & --                & --                & --                & 2 \\ 
2016-03-12 & $57459.47$ & --                & --                & --                & $25.156$($0.105$) & --                & 2 \\
\hline
\end{tabular}
\begin{tablenotes}
\item Numbers in parenthesis correspond to 1\,$\sigma$ statistical uncertainties.
\item  Instrument/Telescope: 1=DECam/Blanco; 2=HSC/SUBARU.
\end{tablenotes}
\end{table*}

\begin{table*}
\caption{$grizy$ photometry of the local sequence stars around UID 30901.} 
\label{tab:sdss}                                                                                                                                                                                                                        
\begin{tabular}{lcccccccc}
\hline
  & RA       &   Dec    & $g$               & $r$               & $i$               & $z$                 & $y$               & SDSS ID$^{\mathrm{a}}$ \\
\hline
 1 & 49.84732 & 2.07194  & $16.825$($0.004$) & $16.430$($0.005$) & $16.290$($0.005$) & $16.253$($0.009$)   & --                & 272 \\
 2 & 49.83740 & 2.08621  & $19.540$($0.013$) & $18.701$($0.010$) & $18.372$($0.010$) & $18.101$($0.028$)   & $18.100$($0.001$) & 377 \\
 3 & 49.83134 & 2.07922  & $21.057$($0.039$) & $20.490$($0.034$) & $20.332$($0.042$) & $20.109$($0.150$)   & $20.250$($0.001$) & 813 \\
 4 & 49.82604 & 2.06618  & $19.358$($0.012$) & $18.941$($0.011$) & $18.759$($0.013$) & $18.788$($0.048$)   & $18.700$($0.001$) & 369 \\
 5 & 49.83355 & 2.05832  & $20.793$($0.031$) & $19.430$($0.015$) & $18.449$($0.011$) & $17.939$($0.025$)   & $17.780$($0.001$) & 431 \\
 6 & 49.83622 & 2.05565  & $22.366$($0.112$) & $21.673$($0.086$) & $21.296$($0.094$) & $21.361$($0.426$)   & $21.130$($0.001$) & 818 \\
 7 & 49.81610 & 2.03973  & $23.981$($0.381$) & $23.161$($0.351$) & $21.209$($0.095$) & $20.706$($0.230$)   & $20.760$($0.002$) & 982 \\
\hline
\end{tabular}
\begin{tablenotes}
\item Numbers in parenthesis correspond to 1\,$\sigma$ statistical uncertainties.
\item $^{\mathrm{a}}$The first 6 stars have SDSS ID prefixed by 1237651753997107. The
7th star SDSS ID is prefixed by 1237653664721928.
\end{tablenotes}
\end{table*}

%


\bsp	
\label{lastpage}
\end{document}